\documentclass[aps, prl, superscriptaddress, twocolumn,
notitlepage, longbibliography,
floatfix]{revtex4-2}

\usepackage{natbib}
\usepackage{graphicx}
\usepackage{mathtools}
\usepackage{amsfonts} 
\usepackage{amsmath} 
\usepackage{amssymb}
\usepackage{amsthm}
\usepackage{bm}
\usepackage{braket}
\usepackage{color} 
\usepackage{bbm}
\usepackage{hyperref}
\usepackage{multirow}
\usepackage{import}
\usepackage{microtype}
\usepackage{relsize}
\usepackage{framed}
\usepackage[dvipsnames]{xcolor}
\usepackage[percent]{overpic}
\usepackage{tikz}
\usepackage{xparse}

\ExplSyntaxOn
\NewDocumentCommand{\colvector}{m}
{
  \begin{pmatrix}
    \clist_use:nn {#1} { \\ }
  \end{pmatrix}
}
\ExplSyntaxOff

\newcommand{\Z}{\mathbb{Z}}

\newtheorem*{corollary*}{Corollary}

\theoremstyle{definition}

\usepackage{newtxtext}
\usepackage{CJK}

\usepackage{graphicx}
\usepackage{dcolumn}
\usepackage{bm}
\usepackage{bbm}
\usepackage{amsthm} 
\usepackage{diagbox}
\usepackage{braket}

\usepackage{amsmath}
\usepackage{graphicx}
\usepackage{enumerate}
\usepackage{epstopdf}
\usepackage{framed}
\usepackage{tabularx}
\usepackage[normalem]{ulem}
\usepackage{booktabs}
\usepackage{float}
\usepackage[dvipsnames]{xcolor}
\usepackage{array}
\usepackage{soul}
\usepackage{tikz}
\hypersetup{colorlinks=true,allcolors=blue}
\usetikzlibrary{positioning}
\usetikzlibrary{decorations,arrows}
\usetikzlibrary{decorations.pathreplacing}
\usetikzlibrary{chains, scopes, positioning, backgrounds, shapes, fit, shadows, calc, arrows.meta}
\tikzset{>=stealth}
\usepackage{color}
\usepackage{wasysym}
\usepackage{mathtools}
\usepackage{comment}
\usetikzlibrary{through,hobby}
\usetikzlibrary{decorations.markings}
\usetikzlibrary{fadings,shadings}
\usetikzlibrary{plotmarks}
\definecolor{darkblue}{rgb}{0.,0.,0.4}
\definecolor{darkred}{rgb}{0.5,0.,0.}
\definecolor{BlueViolet}{RGB}{138,43,226}
\definecolor{SkyBlue}{RGB}{30,144,255}
\definecolor{DarkGreen}{RGB}{0,150,0}
\definecolor{DarkYellow}{RGB}{0,1,1}
\usetikzlibrary{intersections}
\definecolor{iro100}{cmyk}{1,0,0,0}
\definecolor{iro90}{cmyk}{.9,0,0,0}
\definecolor{iro80}{cmyk}{.8,0,0,0}
\definecolor{iro60}{cmyk}{0,.6,0,0}
\definecolor{iro10}{cmyk}{0,.1,0,0}

\newcommand{\1}{\text{\uppercase\expandafter{\romannumeral1}}}
\newcommand{\2}{\text{\uppercase\expandafter{\romannumeral2}}}
\newcommand{\3}{\text{\uppercase\expandafter{\romannumeral3}}}
\newcommand{\4}{\text{\uppercase\expandafter{\romannumeral4}}}
\newcommand{\5}{\text{\uppercase\expandafter{\romannumeral5}}}
\newcommand{\6}{\text{\uppercase\expandafter{\romannumeral6}}}

\def\Z{\mathbb{Z}}

\begin{document}

\author{Wenjie Ji \begin{CJK*}{UTF8}{gkai}(纪文杰)\end{CJK*}}
\affiliation{Perimeter Institute for Theoretical Physics, Waterloo, Ontario, Canada N2L 2Y5}
\affiliation{Department of Physics and Astronomy, McMaster University, Hamilton, Ontario, Canada L8S 4M1}

\author{Ryan A. Lanzetta}
\affiliation{Perimeter Institute for Theoretical Physics, Waterloo, Ontario, Canada N2L 2Y5}

\author{Zheng Zhou (周正)}
\affiliation{Perimeter Institute for Theoretical Physics, Waterloo, Ontario, Canada N2L 2Y5}
\affiliation{Department of Physics and Astronomy, University of Waterloo, Waterloo, Ontario N2L 3G1, Canada}

\author{Chong Wang (王翀)}
\email{cwang4@pitp.ca}
\affiliation{Perimeter Institute for Theoretical Physics, Waterloo, Ontario, Canada N2L 2Y5}

\title{Self-dual Higgs transitions: Toric code and beyond} 

\begin{abstract}
The toric code, when deformed in a way that preserves the self-duality $\mathbb{Z}_2$ symmetry exchanging the electric and magnetic excitations, admits a transition to a topologically trivial state that spontaneously breaks the $\mathbb{Z}_2$ symmetry. Numerically, this transition was found to be continuous, which makes it particularly enigmatic given the longstanding absence of a continuum field-theoretic description. In this work we propose such a continuum field theory for the transition dubbed the $SO(4)_{2,-2}$ Chern-Simons-Higgs (CSH) theory. We show that our field theory provides a natural ``mean-field'' understanding of the phase diagram. Moreover, it can be generalized to an entire series of theories, namely the $SO(4)_{k,-k}$ CSH theories, labeled by an integer $k$. For each $k>2$, the theory describes an analogous transition involving different non-Abelian topological orders, such as the double Fibonacci order ($k=3$) and the $S_3$ quantum double ($k=4$). For $k=1$, we conjecture that the corresponding CSH transition is in fact infrared-dual to the $3d$ Ising transition, in close analogy with the particle-vortex duality of a complex scalar.
 
\end{abstract}

\begin{CJK*}{UTF8}{bkai}
\maketitle
\end{CJK*}


\textit{Introduction. ---}
How does a topological order destroy itself? The problem of quantum phase transitions from topological orders to conventional phases of matter has been a source of inspiration since the early days of topological phases. Consider the $2d$ toric code \cite{Kitaev:1997wr}, or equivalently $(2+1)d$ $\Z_2$ gauge theory \cite{KogutReview}, as a paradigmatic example: by turning on a magnetic field, one can drive the system through a continuous phase transition -- known as the Ising$^*$, or simply ``gauged Ising,'' transition -- from the toric code to a trivial gapped state, via the condensation\footnote{The term ``condensation'' is used in multiple contexts, with its precise meaning depending on the setting. In its most strict sense, it refers to a phase with spontaneous symmetry breaking, where a charged local operator acquires a nonzero expectation value. More broadly, it can also refer to the nonzero expectation values of scalar fields coupled to dynamical gauge fields and is used in the context of Higgs transitions. In addition, in ``anyon condensation,'' it denotes a formal mapping between topological quantum field theories. In this work, we use ``condensation'' in these standard, context-dependent senses; for further clarification and precise terminology, see, e.g., Ref. \onlinecite{cheng2026proliferation}.} of either the electric ($e$) or magnetic ($m$) anyon quasiparticle \cite{Wegner,FradkinShenker}.

\begin{figure}
\centering\includegraphics[width=.4\textwidth]{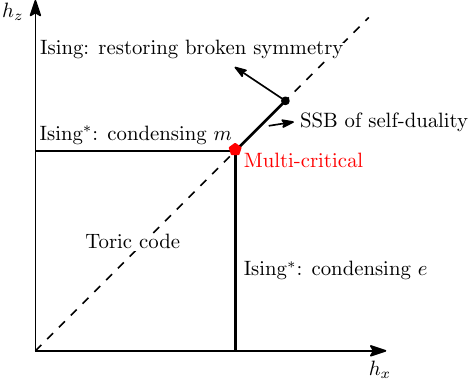}
\caption{Schematic phase diagram of the deformed toric code from numerical studies \cite{TupitsynKitaevProkofEvStamp,VidalDusuelSchmidt,WuDengProkofev,SomozaSernaNahum,OppenheimKochJanuszGazitRingel}. We are interested in the ``multi-critical'' point on the self-dual line $h_x=h_z$, where the two Ising$^*$ transitions meet, followed by a topologically trivial phase that spontaneously breaks the self-duality symmetry.}
\label{Fig:PhaseDiagram}
\end{figure}

The question becomes much more subtle when we impose a $\mathbb{Z}_2$ global symmetry in the toric code that exchanges the $e$ and $m$ quasiparticles -- a symmetry commonly referred to as ``self-duality,'' as it is the direct analogue of electric-magnetic duality in $(3+1)d$ electromagnetism. Because of the mutual statistics between $e$ and $m$ (braiding one around the other yields a $(-1)$ phase), the two particles cannot condense simultaneously. Thus, upon tuning the gaps of both particles to zero at the same time, the resulting transition is expected to be unconventional. Indeed, numerical studies over the past two decades \cite{TupitsynKitaevProkofEvStamp,VidalDusuelSchmidt,WuDengProkofev,SomozaSernaNahum,OppenheimKochJanuszGazitRingel,XuPollmannKnap}, culminating in the first direct evidence for scale invariance in Ref.~\onlinecite{SomozaSernaNahum}, have identified a continuous transition along the self-dual line in parameter space, across which two things occur simultaneously: (1) the topological order disappears, and (2) the $\mathbb{Z}_2$ self-duality symmetry is spontaneously broken. One simple realization of this physics is given by the deformed toric code Hamiltonian~\cite{TupitsynKitaevProkofEvStamp} 
\begin{equation}
\label{eq:deformedTC}
    H=H_{\text{Toric-code}}-h_x\sum_e{\sigma^x_e}-h_z\sum_e \sigma^z_e
\end{equation}
Schematically, the phase diagram is shown in Fig.~\ref{Fig:PhaseDiagram} \footnote{Actual numerical simulations often work with Euclidean lattice actions instead of Hamiltonians. The universal physics is expected to be the same.}. 

The numerical findings pose a natural question for theorists: is there a continuum field-theoretic description of this self-dual transition of the toric code? Despite the simple and natural form of the deformed toric code Hamiltonian, Eq.~\eqref{eq:deformedTC}, a satisfactory field theory formulation has so far not been discovered. The difficulty lies in formulating a field theory that both (a) contains degrees of freedom with mutual statistics, and (b) exhibits spontaneous-symmetry-breaking (SSB) order on the non-topological side in a natural way. These two requirements can be summarized as the need to ``produce the correct phase diagram.'' We note that several field theories have been proposed in the literature. We briefly review two such existing proposals, and explain the issues with the proposed theories in satisfying both criteria listed above in the Supplementary Material\cite{supp}.

In this work we propose that a Chern-Simons-Higgs (CSH) continuum field theory, more specifically the $SO(4)_{2,-2}$ CSH theory, naturally describes the self-dual transition of the toric code. A helpful picture is to consider a parent double-Ising topological order, in which the $e$ and $m$ particles are unified into a single non-Abelian anyon $e\oplus m\sim \sigma\bar{\sigma}$. From this point of view, the competing condensations of $e$ and $m$ can be understood as the condensation of the non-Abelian anyon $\sigma\bar{\sigma}$. Besides providing a ``mean-field'' understanding of this otherwise mysterious phase transition, our framework also \textit{predicts} an entire series of analogous transitions involving various non-Abelian topological orders -- like the $S_3$ quantum double -- by modifying the Chern-Simons (CS) level. In particular, mean-field analysis of the $SO(4)_{1,-1}$ CSH theory leads us to conjecture that, surprisingly, it is dual to the familiar $3d$ Ising criticality.

\textit{The $SO(4)_{2,-2}$ CSH field theory. ---}
We claim that the following $(2+1)d$ field theory describes the self-dual transition of the toric code:
\begin{multline}
    S=\int d^3x\left[(D_{\mathcal{A}}\Phi)^2+r\Phi^2+\lambda\Phi^4+\frac{1}{g}\operatorname{tr}\mathcal{F_{\mu\nu}}^2\right]\\+i{\rm CS}[\mathcal{A}]_{2,-2}
    \label{eq:2,-2Higgs}
\end{multline}
Here, $\mathcal{A}$ is a dynamical $SO(4)$ gauge field, $\Phi$ is a $4$-component real scalar that transforms as a vector under the $SO(4)$ gauge symmetry. The above theory is essentially a Higgs transition of $SO(4)$ gauge theory with the standard $\Phi^4$ and Yang-Mills $\operatorname{tr}\mathcal{F}_{\mu\nu}^2$ terms, except with a specific CS term $\rm CS[\mathcal{A}]_{2,-2}$. Since $SO(4)=\frac{SU(2)_L\times SU(2)_R}{\mathbb{Z}_2}$, there are two independent CS terms from each $SU(2)$, the notation $(2,-2)$ means that we have level $2$ for $SU(2)_L$ and level $-2$ for $SU(2)_R$. More explicitly, let $\mathcal{A}_{L,R}$ be the gauge field corresponding to the $SU(2)_{L,R}$ Lie algebra, the $SO(4)_{2,-2}$ Chern-Simons term reads
\begin{multline}
    {\rm CS}[\mathcal{A}]_{2,-2}=\frac{2}{4\pi}\int {\rm tr}\left(\mathcal{A}_L\wedge d\mathcal{A}_L+\frac{2}{3}\mathcal{A}_L\wedge\mathcal{A}_L\wedge\mathcal{A}_L \right) \nonumber \\
    -\frac{2}{4\pi}\int {\rm tr}\left(\mathcal{A}_R\wedge d\mathcal{A}_R+\frac{2}{3}\mathcal{A}_R\wedge\mathcal{A}_R\wedge\mathcal{A}_R \right). \nonumber
\end{multline}

We now discuss global symmetries of Eq.~\eqref{eq:2,-2Higgs}. First, there is a $\mathbb{Z}_2^T$ symmetry that simultaneously performs time reversal and exchanges $SU(2)_L\leftrightarrow SU(2)_R$, so that the CS term remains invariant under time-reversal. This means that time-reversal should act on the $\Phi$ scalar as an improper $\Z_2$, for example as $(\Phi_1,\Phi_2,\Phi_3,\Phi_4)\to (-\Phi_1,\Phi_2,\Phi_3,\Phi_4)$ up to gauge transformations. There is also a unitary $\Z_2$ symmetry for the $SO(4)$ gauge theory. Recall that $\pi_1(SO(4))=\Z_2$, so on any closed $2$-dimensional surface there is a $\Z_2$ gauge flux that is conserved. A concrete realization of a nontrivial flux is to pick a generator of $SO(2)\subset SO(4)$ and insert a $2\pi$ flux quantum of this $SO(2)$. The conserved gauge flux corresponds to a global zero-form $\Z_2$ symmetry, and the corresponding charged operators are (by definition) the monopole operators of the $SO(4)$ gauge field. The conserved $\Z_2$ flux through a $2$-surface $\mathcal{C}_2$ is measured by the second Stiefel-Whitney class  of the $SO(4)$ gauge field $\int_{\mathcal{C}_2}w_2^{\mathcal{A}}\in\{0,1\}$ -- a direct $\Z_2$ analogue of the integer magnetic flux for $U(1)$ gauge fields $\int dA$. A compact way to specify this $\Z_2$ flux symmetry is to introduce a background $\Z_2$ gauge field $A^{\Z_2}$, which couples to the theory through a topological term
\begin{equation}
    \label{eq:topocoupling}
    i\pi\int A^{\Z_2}\cup w_2^{\mathcal{A}},
\end{equation}
where $\cup$ is the standard cup product for discrete cohomology. The above term is a direct $\Z_2$ analogue of the mutual Chern-Simons coupling $A\wedge da$ in $U(1)$ gauge theories.

Let us now verify that the theory in Eq.~\eqref{eq:2,-2Higgs} indeed produces the correct phase diagram. In particular, we will show that for $r>0$ the theory realizes the $\Z_2$ topological order, with the flux $\Z_2$ symmetry acting as the $e\leftrightarrow m$ self-duality symmetry; and for $r<0$ the topological order is lost while the $\Z_2$ symmetry is spontaneously broken.

\textbf{\boldmath The topological side: $r>0$.} The scalar field $\Phi$ is gapped and can be safely integrated out. At low energy we get a pure $SO(4)_{2,-2}$ Chern-Simons theory, which is really a $\Z_2$ gauge theory in disguise. To see this, we first pretend that the group is not $SO(4)$, but instead $Spin(4)=SU(2)_L\times SU(2)_R$. Then we just have two decoupled CS theories: $SU(2)_2\times SU(2)_{-2}$. 

Recall that $SU(2)_2$ CS theory describes an Ising-like topological order, with anyons $\{1,\sigma,\psi\}$, fusion rules $\psi\times \psi= 1$, $\psi\times \sigma= \sigma$ and $\sigma\times\sigma= 1+\psi$. The $\psi$ particle is a fermion with topological spin $\theta_{\psi}=-1$, and the Ising-like anyon $\sigma$ has $\theta_{\sigma}=e^{i3\pi/8}$. The only difference from the standard Ising topological order is that the Ising topological order has $\theta_{\sigma}=e^{i\pi/8}$. In terms of Kitaev's $16$-fold classification~\cite{Kitaev:2005hzj}, standard Ising corresponds to $\nu=1$ (mod $16$) while $SU(2)_2$ corresponds to $\nu=3$ (mod $16$). 

Now we go back to the $SU(2)_2\times SU(2)_{-2}$ theory, with anyon contents $\{1,\sigma,\psi\}\times \{1,\bar{\sigma},\bar{\psi}\}$, where $\theta_{\sigma}=e^{i3\pi/8}$ and $\theta_{\bar{\sigma}}=e^{-i3\pi/8}$. The effect of modding out the common $\Z_2$ center in the gauge group (which makes it $SO(4)$ instead of $Spin(4)$) is to ``glue'' $\sigma$ and $\bar{\sigma}$ together, which confines unpaired $\sigma$ and $\bar{\sigma}$ anyons from the excitation spectrum.  Equivalently, we have condensed the combination $\psi\bar{\psi}$, which in field theory language can be stated as gauging the $\Z_2$ one-form symmetry generated by the $\psi\bar{\psi}$ topological line \cite{BhardwajGaiottoKapustin2017}. The resulting state is well known to be nothing but the toric code, with $\psi\sim\bar{\psi}$ as the fermionic excitation, and $\sigma\bar{\sigma}=e\oplus m$. More details on this anyon condensation can be found in Ref.~\cite{bais2009condensate} and Supplementary Materials.

Next we verify that the $\Z_2$ flux symmetry indeed acts as the self-duality symmetry exchanging the $e$ and $m$ particles in the toric code. For this purpose, consider gauging the $\Z_2$ symmetry, by promoting $A_{\Z_2}$ in Eq.~\eqref{eq:topocoupling} to a dynamical gauge field, which we denote as $a$. Since the topological coupling Eq.~\eqref{eq:topocoupling} is the only term involving $a$, the sole effect of integrating out $a$ is to set $w_2^{SO(4)}=0$. Recall that the second Stiefel-Whitney class serves as the obstruction to lift the $SO(n)$ gauge field to a $Spin(n)$ gauge field. So the vanishing of $w_2$ means that we are now actually dealing with a true $Spin(4)=SU(2)\times SU(2)$ gauge theory, which is nothing but the $SU(2)_2\times SU(2)_{-2}$ CS theory.

Indeed, the $SU(2)_2\times SU(2)_{-2}$ ``double-Ising'' topological order is what we expect from gauging the $\Z_2$ self-duality symmetry of the toric code, where the $\Z_2$ gauge flux becomes non-Abelian ($\sigma$ or $\bar{\sigma}$). If the $\Z_2$ symmetry acts on the toric code in other ways (without exchanging $e$ and $m$), then gauging it would not result in the non-Abelian ``double-Ising'' topological order \cite{TeoHughesFradkin}. We have therefore verified that the $\Z_2$ flux symmetry acts as the self-duality symmetry in the toric code phase. In fact, in the Supplementary Materials we show that one appealing feature of the $SO(4)_{2,-2}$ representation of the toric code is its manifest $\mathbb{Z}_2 \times \mathbb{Z}_2^T$ symmetry, in sharp contrast to the more commonly used Abelian mutual Chern-Simons description.

Now we comment more about gauging the $\Z_2$ self-duality symmetry of toric code: in general we should expect the anyon contents to be $\{1,\sigma,\psi\}\times \{1,\bar{\sigma},\bar{\psi}\}$, with the spin of the Ising-like anyon $\theta_{\sigma}=e^{i\nu\pi/8}$ ($\nu$ an odd integer \cite{Kitaev:2005hzj}). There are two inequivalent choices: $\nu=1$, which is the standard double-Ising, or $\nu=3$ which is $SU(2)_2\times SU(2)_{-2}$ \footnote{Naively there are more: $\nu$ can take any odd integer mod $16$ \cite{Kitaev:2005hzj}. Obviously $\nu$ is equivalent to $-\nu$ since the theory is doubled. It turns out that $\nu$ is also equivalent to $\nu+8$ through the following anyon relabeling: $\sigma'=\sigma\times\bar{\psi}$ and $\bar{\sigma}'=\bar{\sigma}\times \psi$. Then we are left with only two distinct possibilities.}. As we explained above, gauging the $\Z_2$ flux symmetry from Eq.~\eqref{eq:2,-2Higgs} and \eqref{eq:topocoupling} leads to the $\nu=3$ theory. To obtain the $\nu=1$ theory, we can add another topological term
\begin{equation}
\label{eq:DWterm}
    i\pi \int A^{\Z_2}\cup A^{\Z_2}\cup A^{\Z_2}.
\end{equation}
Physically, this stacks a symmetry-protected topological (SPT) state of the $\Z_2$ symmetry (also known as the Levin-Gu state \cite{LevinGu}). The effect of the SPT stacking is to change the topological spin of the $\Z_2$ flux by $\pm i$ -- which converts $\theta_{\sigma}=e^{\pm i3\pi/8}$ to $e^{\mp i\pi/8}$. Critical properties like scaling dimensions are, however, unaffected by such a topological background term.

So far we have discussed two different gauging processes: (1) gauging the one-form $\Z_2$ symmetry, corresponding to the $\psi\bar{\psi}$ line, from the $SU(2)_2\times SU(2)_{-2}$ topological order {results in the toric code}; (2) gauging the zero-form $\Z_2$ self-duality symmetry from the toric code {results in the $SU(2)_2\times SU(2)_{-2}$ topological order}.
As shown from our discussions, the two processes are in fact the inverse of each other: performing one gauging after the other brings us back to the starting theory. This is a general phenomenon: gauging a one-form (zero-form) $\Z_2$ symmetry in $(2+1)d$ leads to a global zero-form (one-form) $\Z_2$ symmetry, with the $\Z_2$ symmetry generated by $\int_\mathcal{C} b$, where $b$ is the dynamical two-form (one-form) $\Z_2$ gauge field, and $\mathcal{C}$ is any two-cycle (one-cycle) in spacetime \cite{HigherFormSym}.

\textbf{\boldmath The symmetry-breaking side: $r<0$.} The field $\Phi$ now condenses, $\langle\Phi\rangle\neq0$. As $\Phi$ is an $SO(4)$ vector, the $SO(4)$ gauge symmetry is Higgsed down to $SO(3)$, which is the ``diagonal'' $SU(2)$, i.e. identical rotations in $SU(2)_L$ and $SU(2)_R$. This means that the Higgs condensate sets $\mathcal{A}_L=\mathcal{A}_R$. As a result, the two Chern-Simons terms, one at level $+2$ and the other at level $-2$, cancel each other, so we are left with a pure Yang-Mills theory of the $SO(3)=SU(2)/\mathbb{Z}_2$ gauge field. As we shall explain below, at low energy the $SO(3)$ Yang-Mills theory in $(2+1)d$ describes a gapped phase without topological order, but with spontaneously broken $\Z_2$ flux symmetry  \cite{KovnerReview,ReadingBetweenLines}. 

Again consider gauging the $\mathbb{Z}_2$ flux symmetry of the $SO(3)$ gauge field. Integrating out the $\mathbb{Z}_2$ gauge field in Eq.~\eqref{eq:topocoupling} once again imposes the constraint $w_2^{SO(3)} = 0$. Since $w_2^{SO(3)}$ is precisely the obstruction to lifting an $SO(3)$ gauge bundle to an $SU(2)$ one, enforcing that $w_2^{SO(3)}=0$ yields an $SU(2)$ Yang-Mills theory. It is quite reasonable to assume -- supported amply by numerical evidence \cite{Teper,AthenodorouTeper} -- that $SU(2)$ Yang-Mills confines at low energies, with a unique gapped vacuum (ground state). But this in turn implies that the theory before gauging the $\mathbb{Z}_2$, namely the $SO(3)$ Yang-Mills theory, must spontaneously break its $\mathbb{Z}_2$ symmetry. Equivalently, we may start from $SU(2)$ Yang-Mills and obtain $SO(3)$ Yang-Mills by gauging the $\mathbb{Z}_2$ one-form symmetry of the former. Since $SU(2)$ Yang-Mills has a trivial vacuum, gauging its one-form symmetry necessarily leads to the spontaneous breaking of the corresponding zero-form $\mathbb{Z}_2$ symmetry.

This completes the matching of the phase diagram.

\textbf{\boldmath Critical point: $r=0$.} More precisely, the critical point is reached when $r=r_c$, for some $r_c$ that depends on other couplings $\lambda$ and $g$ in Eq.~\eqref{eq:2,-2Higgs}. The Higgs field theory Eq.~\eqref{eq:2,-2Higgs} is super-renormalizable in $(2+1)d$, with both the $\Phi^4$-coupling $\lambda$ and Yang-Mills gauge coupling $g$ carrying dimensions of energy. This implies that the theory is free at high energies and therefore well-defined as a continuum field theory. As in other similar critical gauge theories in $(2+1)d$, achieving a continuous transition (assuming one exists) requires $\lambda\gg g$. In other words, as the system flows toward the IR, it should first approach the $O(4)$ Wilson-Fisher fixed point before the gauge coupling becomes important \cite{DasguptaHalperin}.

At the critical point, the theory flows to strong coupling in the IR and we do not have analytic control of its critical properties. In particular, gauge fluctuation may drive the transition to a first-order one~\cite{Coleman1973,Halperin1974}, a scenario difficult to rule out analytically. We can generalize the theory to an $SO(4)_{k,-k}$ CSH theory for arbitrary integer $k$. If the Chern-Simons level $k$ is sufficiently large, the gauge fluctuation will be suppressed, and some critical exponent, such as the scaling dimension of the boson mass operator $\Delta_{\Phi^2}=3-1/\nu$ will approach that of the $O(4)$ Wilson-Fisher theory, which is known quite accurately through numerical bootstrap \cite{O(N)Bootstrap} and Monte Carlo simulation \cite{O(4)MonteCarlo}. Since the theory is invariant under $k\to-k$, we expect
\begin{equation}
\label{eq:largek}
    \nu_k=\nu_{O(4){\rm~WF}}-O\left(\frac{1}{k^2}\right)\approx 0.75-O\left(\frac{1}{k^2}\right).
\end{equation}
The current numerical estimation sits between $0.65$--$0.69$ \cite{SomozaSernaNahum,OppenheimKochJanuszGazitRingel}, which is encouraging although further studies are certainly required to better confirm the picture. Perhaps a more interesting operator is the $\Z_2$-odd monopole. We expect the scaling dimension of the monopole operator to grow with $k$. For sufficiently large $k$, the monopole operator will become irrelevant, implying that the $\Z_2$ symmetry will emerge at the critical point even if microscopically absent. We shall leave the systematic expansion (in large $k$, for example) to future works.

Starting from the critical theory at $r_c$, we can explicitly break the $\mathbb{Z}_2$ symmetry by turning on the $\mathbb{Z}_2$-odd monopole operator. In light of the full phase diagram in Fig.~\ref{Fig:PhaseDiagram}, we expect this perturbation to drive the theory toward the $(2+1)d$ Ising$^*$ universality class~\footnote{More accurately, one must additionally tune the mass term $\Phi^2$ to reach the Ising$^*$ critical point. The shape of the transition line is then controlled by the scaling dimensions of the mass and monopole operators.}. Providing an explicit demonstration of such a renormalization-group flow at strong coupling is, however, generally a difficult problem. A closely related situation arises in the celebrated deconfined critical point \cite{deconfine1,deconfine2}, described by the $\mathbb{CP}^1$ Abelian Higgs theory. There, physical considerations strongly suggest that turning on a monopole operator drives the theory to the ordinary $O(3)$ Wilson-Fisher critical point, yet no explicit analytic derivation of this flow is currently available.

As discussed above, if we gauge the global $\Z_2$ symmetry throughout the transition, we obtain the closely related $SU(2)_2\times SU(2)_{-2}$ CSH theory, describing a transition from a doubled $SU(2)_2$ topological order, or a double-Ising if we also add Eq.~\eqref{eq:DWterm}, to a trivial confined phase. It is known that these ``double Ising'' type of topological orders can be confined through anyon condensation of $\sigma\bar{\sigma}$ and $\psi\bar{\psi}$. Since $\sigma$ ($\bar{\sigma}$) is represented in the Chern-Simons theory as the fundamental representation of $SU(2)$, our scalar field $\Phi$ can be interpreted in the topological phase as the combined particle $\sigma\bar{\sigma}$. Our field theory can therefore be interpreted as a continuous transition involving the condensation of $\sigma\bar{\sigma}$. The other particle $\psi\bar{\psi}$ is automatically condensed as it appears in the fusion product of $\sigma\bar{\sigma}$ with itself. We note that, in general, a Chern-Simons-Higgs theory does not necessarily correspond to any anyon condensation. The simplest example is the transition from a Laughlin state to a trivial state, which can be described by a $U(1)_k$ CSH theory; yet this transition clearly does not involve anyon condensation, since there is no bosonic anyon available to condense in the first place. In our case, however, the Higgs picture does neatly correspond to the the $\sigma\bar{\sigma}$ condensation.

\textit{Generalization: $SO(4)_{k,-k}$  CSH theory. ---}
We can generalize Eq.~\eqref{eq:2,-2Higgs} to other values of $k\in\Z$. As already discussed, for sufficiently large $k$ the gauge fluctuation will be controlled and we expect results like Eq.~\eqref{eq:largek}. In this Section we show that the theory at different values of $k$ correspond to interesting phase transitions involving different non-Abelian topological orders.

For any value of $k>0$, we can follow exactly the same steps as before to conclude that the field theory describes a transition from a topological order, described by the $SO(4)_{k,-k}$ Chern-Simons theory, to a trivial topological order with a spontaneously broken global $\Z_2$ symmetry. The action of the $\Z_2$ symmetry will be different for even and odd values of $k$. For even values of $k$, the $\Z_2$ symmetry will exchange some anyons, similar to the $e\leftrightarrow m$ duality in the toric code.  For odd values of $k$, however, the $\Z_2$ symmetry will act trivially on the topological order -- we then have a peculiar situation, where the topological transition and the symmetry-breaking transition happen at the point, even though the topological order and the symmetry decouple throughout. In the End Matter we include more details on the $\Z_2$ symmetry action.

We now illustrate with some simple examples.

\textbf{\boldmath $k=4$: $S_3$ quantum double.} The $SO(4)_{4,-4}$ CS theory describes the untwisted $S_3$ quantum double $\mathcal{D}(S_3)$ \cite{bais2009condensate}. The global $\Z_2$ symmetry from the $SO(4)$ flux conservation acts as a nontrivial anyon automorphism in $\mathcal{D}(S_3)$: it exchanges the $d=2$ flux with the $d=2$ charge ($d$ being the quantum dimension). The fact that the $\Z_2$ symmetry acts as a nontrivial anyon-permuting symmetry is a general feature for all even $k$. We have thus made a rather nontrivial prediction, namely there is a direct, continuous transition from an $S_3$ quantum double, with a nontrivial $\Z_2$ anyon permutation symmetry, to a topologically trivial state that spontaneously breaks the $\Z_2$ symmetry. It will be interesting to realize this transition in future numerical simulations.

Let us now better understand the transition from the anyon condensation perspective. In the $SU(2)_4\times SU(2)_{-4}$ CS theory with anyon content $\{0,\frac{1}{2},1,\frac{3}{2},2\}\times \{0,\frac{1}{2},1,\frac{3}{2},2\}$, there is a condensable algebra $(0,0) + (2,2)$. Here, $(2,2)$ is an abelian boson with $\mathbb{Z}_2$ fusion rule. When $(2,2)$ condenses, the phase becomes the $S_3$ quantum double, or rather, the $SO(4)_{4,-4}$ CS theory. In the meantime, the non-Abelian anyon $(1,1)$ with quantum dimension $4$ splits into the $d=2$ charge and the $d=2$ flux in $\mathcal{D}(S_3)$. In terms of anyons in $\mathcal{D}(S_3)$ (shown in Table \ref{fig:S3_anyons}), the remaining deconfined anyons are
\begin{equation}
\begin{aligned}
(0,0)\sim (2,2) &\rightarrow A \text{ (identity)}&
(0,2)\sim (2,0) & \rightarrow B\\
(1,1) & \rightarrow C + F \\
\left(\tfrac{1}{2},\tfrac{1}{2}\right)\sim \left(\tfrac{3}{2},\tfrac{3}{2}\right) &\rightarrow D&
\left(\tfrac{1}{2},\tfrac{3}{2}\right)\sim \left(\tfrac{3}{2},\tfrac{1}{2}\right) &\rightarrow E\\
(1,0)\sim (1,2) & \rightarrow G&
(0,1)\sim (2,1) & \rightarrow H.
\end{aligned}
\end{equation}

\begin{table}
\centering
\renewcommand{\arraystretch}{1.2}
\begin{tabular}{| c | c c c | c c | c c c |}
\hline
 & $ A $ & $ B $ & $ C $ & $ D $ & $ E $ & $ F $ & $ G $ & $ H $ \\
 \hline
conjugacy class & $e$ & $e$ & $e$ & $s$ & $s$ & $r$ &  $r$ & $r$ \\
irrep of the centralizer & $1$ & sign & $[2]$ & $1$ & $[-1]$ & $1$ & $[\omega]$ & $[\bar{\omega}]$ \\
\hline
quantum dimension & 1 & 1 & 2 & 3 & 3 & 2 & 2 & 2 \\
\hline
topological spin \rule{0pt}{3ex} & 1 & 1 & 1 & 1 & $-1$ & 1 & $e^{i\frac{2\pi}{3}}$ & $e^{-i\frac{2\pi}{3}}$ \\
\hline
\end{tabular}
\caption{Anyons and their quantum dimension in $\mathcal{D}(S_3)$  \cite{Kitaev:1997wr,beigi2011quantum}. $S_3=\langle s,r: s^2=r^3=e,srs = r^{-1}\rangle$. Here, sign, $[2]$ denote the sign rep and 2-dimensional rep of $S_3$; $[-1]$ is the non-trivial rep of $\mathbb{Z}_2$; $[\omega]$ and $[\bar{\omega}]$ denote the two non-trivial rep of $\mathbb{Z}_3$ that are conjugate to each other.}
\label{fig:S3_anyons}
\end{table}

In the $SO(4)_{4,-4}$ CS theory, when the $4$-component real scalar $\Phi$ (the $(\frac{1}{2},\frac{1}{2})$ in terms of $SU(2)_4\times SU(2)_{-4}$ labeling) condenses, so are its bosonic fusion outcomes. Correspondingly, in $\mathcal{D}(S_3)$, condensable algebras that include $D$ are either $A+C+D$ or $A+F+D$, both are Langrangian algebras. Therefore, when $D$ condenses, the phase has trivial topological order. Since meanwhile, the $\mathbb{Z}_2$ symmetry that exchanges $C$ and $F$ is spontaneously broken, the resulting phase is the first-order transition line between the $A+C+D$ condensate and the $A+F+D$ condensate.

\textbf{\boldmath $k=3$: double Fibonacci.} The $SO(4)_{3,-3}$ CS theory describes the double Fibonacci topological order $\{1,\tau\}\times\{1,\bar{\tau}\}$. So the $k=3$ Higgs theory describes a transition from double Fibonacci to a trivial topological order, again accompanied with the spontaneous breaking of a global $\Z_2$ symmetry. Curiously, unlike the previous case, this global $\Z_2$ symmetry acts trivially in the double Fibonacci topological order -- the simplest way to see this is to notice that there is no way to nontrivially enrich the double Fibonacci with a global $\Z_2$ symmetry. So topologically speaking, the $\Z_2$ symmetry and double Fibonacci are ``decoupled'' away from the critical point. This is a general feature for all odd $k$.

\textbf{\boldmath $k=1$: the Ising transition.} We can also consider the $k=1$ version of the $SO(4)_{k,-k}$ CSH transition. The Higgs phase, for the same reason as before, is a topologically trivial phase with spontaneously broken $\Z_2$ global symmetry. The symmetric phase, described by $SO(4)_{1,-1}$ CS theory, is in fact a trivial topological order, with no nontrivial anyon excitation at all. To see this, we start from $SU(2)_1\times SU(2)_{-1}$, which is an abelian double-semion topological order, with anyons $\{1,s,\bar{s},b=s\bar{s}\}$ where $s$ ($\bar{s}$) is a semion (anti-semion) with topological spin $\theta_s=i$ ($\theta_{\bar{s}}=-i$). To recover $SO(4)_{1,-1}=SU(2)_1\times SU(2)_{-1}/\Z_2$, we need to condense the $b=s\bar{s}$ boson, which completely confines the topological order since both $s$ and $\bar{s}$ have non-trivial mutual  braiding statistics $(-1)$ with $b$. 

The theory thus describes a transition from a trivial vacuum to a $\Z_2$ ferromagnet. A natural conjecture is that this is simply a dual description of the $3d$ Ising transition. (The other possibility is that our gauge theory describes a first-order transition.) In the dual description, the Ising spin $\phi$ corresponds to the $\Z_2$ monopole operator $\mathcal{M}_{SO(4)}$, in direct analogy with the particle-vortex duality of the XY model in $3d$ and the Kramers-Wannier duality of the Ising model in $2d$ \cite{DualityReview}.

\textit{Discussions. ---}
In this work we have demystified the numerically observed self-dual transition of the toric code by proposing that an $SO(4)_{2,-2}$ CSH field theory naturally captures the observed phase diagram: a toric-code phase with a self-duality $\mathbb{Z}_2$ symmetry on one side, and a topologically trivial phase with spontaneous $\mathbb{Z}_2$ symmetry breaking on the other. This framework extends straightforwardly and naturally predicts an entire series of exotic transitions involving non-Abelian topological orders such as the $S_3$ quantum double, where an anyon-permuting symmetry is spontaneously broken on the non-topological side. 

Conceptually, our field theory shares some key features with the familiar story of ``deconfined criticality,'' where the system is embedded into a gauge theory with a larger gauge group to produce an unconventional phase diagram, and where spontaneous symmetry breaking arises through instanton proliferation in the gauge theory \cite{deconfine1,deconfine2}. However, there is an important difference: the standard deconfined critical point (and its close relatives) exhibits an enlarged global symmetry in the IR -- such as the well-known $SO(5)$ at the Neel to valence-bond-solid transition -- and this IR symmetry carries a nontrivial 't Hooft anomaly \cite{SO(5),SO5}. In contrast, our $SO(4)_{k,-k}$ CSH theory does not display any manifest enlarged IR symmetry. In fact, an interesting feature of our theory is the apparent \textit{absence} of additional IR symmetries, either zero-form or one-form, beyond the microscopic $\Z_2^T\times \Z_2$. This is also consistent with recent numerical study that finds no additional local, conserved current density \cite{OppenheimKochJanuszGazitRingel}. \par 
The lack of additional IR symmetries at the multicritical point also poses an interesting challenge to isolating this theory with the conformal bootstrap. Existing numerical studies suggest, but do not conclusively demonstrate, that the theory contains only two non-trivial relevant local scalar operators, one of which is $\mathbb Z_2$-even and the other $\mathbb Z_2$-odd, which is the case also for the $3d$ Ising CFT.  Taking this literally, the multicritical theory sits within the ``continent" of the bootstrap allowed region for this class of CFTs \cite{Kos:2014bka}, a parameter regime where additional input is presumably required to yield precision results analogous to $3d$ Ising \cite{El-Showk:2012cjh,Kos:2014bka,Chang:2024whx}. One observation is that the self-dual multicritical point of toric code does not contain any obvious ``pinning"-type line defects created by coupling a local operator with $\Delta < 1$ to a line, since the measured scaling dimensions of local operators satisfy $\Delta > 1$ \cite{SomozaSernaNahum}, in stark contrast to the $3d$ Ising CFT whose $\mathbb Z_2$-breaking pinning line defect has been studied extensively \cite{allais2014magnetic,Allais:2014fqa,Parisen_Toldin_2017,Cuomo:2021kfm,Gimenez-Grau:2022ebb,BianchiBootstrapMagneticField,Nishioka:2022qmj,Hu_2024,Zhou:2023fqu,Lanzetta:2025xfw}. We expect this observation could be important in an attempt to bootstrap the theory following e.g. \cite{El-Showk:2012cjh,Kos:2014bka,Chang:2024whx,Lanzetta:2025xfw}, although  we leave the issue of whether the multicritical theory supports \emph{any} IR-stable genuine line defect as an interesting open problem. \par 
A final general lesson from our study is that when contemplating the ``condensation'' of multiple anyons that are mutually non-local -- and therefore cannot truly condense simultaneously -- it is useful to introduce a ``parent'' non-Abelian topological order in which these anyons are ``unified'' into a single non-Abelian excitation. In the toric-code example, the relevant parent order is the double Ising theory, with the identification $e\oplus m\sim \sigma\bar{\sigma}$. Viewed this way, our work provides a concrete physical application, in the setting of continuous quantum phase transitions, of the abstract notion of non-Abelian anyon condensation. Armed with this understanding, many other exotic transitions involving non-Abelian topological orders -- such as the $S_3$ quantum double with its anyon-permuting symmetry -- become natural possibilities. It would be very interesting to identify a numerical realization of such non-Abelian transitions in the near future. \par

\textbf{Acknowledgments}: We thank Davide Gaiotto, Jaume Gomis, Tarun Grover, John McGreevy, David Poland, and Nathan Seiberg for insightful discussions. CW would like to especially thank Yin-Chen He, Adam Nahum, and Weicheng Ye for many conversations on this problem over the past few years. Research at Perimeter Institute is supported in part by the Government of Canada through the Department of Innovation, Science and Industry Canada and by the Province of Ontario through the Ministry of Colleges and Universities. We also acknowledge support from the Natural Sciences and Engineering Research Council of Canada (NSERC) through Discovery Grant awards (WJ and ZZ) and a Discovery Launch Supplement award (WJ). Part of this work was done at the Kavli Institute for Theoretical Physics during the ``Noise-robust Phases of Quantum Matter'' program, which was supported in part by grant NSF PHY-2309135 to the Kavli Institute for Theoretical Physics (KITP).

\bibliography{References}

\providecommand{\noopsort}[1]{}\providecommand{\singleletter}[1]{#1}%
\begin{thebibliography}{63}%
\makeatletter
\providecommand \@ifxundefined [1]{%
 \@ifx{#1\undefined}
}%
\providecommand \@ifnum [1]{%
 \ifnum #1\expandafter \@firstoftwo
 \else \expandafter \@secondoftwo
 \fi
}%
\providecommand \@ifx [1]{%
 \ifx #1\expandafter \@firstoftwo
 \else \expandafter \@secondoftwo
 \fi
}%
\providecommand \natexlab [1]{#1}%
\providecommand \enquote  [1]{``#1''}%
\providecommand \bibnamefont  [1]{#1}%
\providecommand \bibfnamefont [1]{#1}%
\providecommand \citenamefont [1]{#1}%
\providecommand \href@noop [0]{\@secondoftwo}%
\providecommand \href [0]{\begingroup \@sanitize@url \@href}%
\providecommand \@href[1]{\@@startlink{#1}\@@href}%
\providecommand \@@href[1]{\endgroup#1\@@endlink}%
\providecommand \@sanitize@url [0]{\catcode `\\12\catcode `\$12\catcode
  `\&12\catcode `\#12\catcode `\^12\catcode `\_12\catcode `\%12\relax}%
\providecommand \@@startlink[1]{}%
\providecommand \@@endlink[0]{}%
\providecommand \url  [0]{\begingroup\@sanitize@url \@url }%
\providecommand \@url [1]{\endgroup\@href {#1}{\urlprefix }}%
\providecommand \urlprefix  [0]{URL }%
\providecommand \Eprint [0]{\href }%
\providecommand \doibase [0]{https://doi.org/}%
\providecommand \selectlanguage [0]{\@gobble}%
\providecommand \bibinfo  [0]{\@secondoftwo}%
\providecommand \bibfield  [0]{\@secondoftwo}%
\providecommand \translation [1]{[#1]}%
\providecommand \BibitemOpen [0]{}%
\providecommand \bibitemStop [0]{}%
\providecommand \bibitemNoStop [0]{.\EOS\space}%
\providecommand \EOS [0]{\spacefactor3000\relax}%
\providecommand \BibitemShut  [1]{\csname bibitem#1\endcsname}%
\let\auto@bib@innerbib\@empty
\bibitem [{\citenamefont {Kitaev}(2003)}]{Kitaev:1997wr}%
  \BibitemOpen
  \bibfield  {author} {\bibinfo {author} {\bibfnamefont {A.~Y.}\ \bibnamefont
  {Kitaev}},\ }\bibfield  {title} {\bibinfo {title} {Fault-tolerant quantum
  computation by anyons},\ }\href
  {https://doi.org/10.1016/S0003-4916(02)00018-0} {\bibfield  {journal}
  {\bibinfo  {journal} {Ann. Phys.}\ }\textbf {\bibinfo {volume} {303}},\
  \bibinfo {pages} {2} (\bibinfo {year} {2003})},\ \Eprint
  {https://arxiv.org/abs/quant-ph/9707021} {arXiv:quant-ph/9707021}
  \BibitemShut {NoStop}%
\bibitem [{\citenamefont {Kogut}(1979)}]{KogutReview}%
  \BibitemOpen
  \bibfield  {author} {\bibinfo {author} {\bibfnamefont {J.~B.}\ \bibnamefont
  {Kogut}},\ }\bibfield  {title} {\bibinfo {title} {An introduction to lattice
  gauge theory and spin systems},\ }\href
  {https://doi.org/10.1103/RevModPhys.51.659} {\bibfield  {journal} {\bibinfo
  {journal} {Rev. Mod. Phys.}\ }\textbf {\bibinfo {volume} {51}},\ \bibinfo
  {pages} {659} (\bibinfo {year} {1979})}\BibitemShut {NoStop}%
\bibitem [{Note1()}]{Note1}%
  \BibitemOpen
  \bibinfo {note} {The term ``condensation'' is used in multiple contexts, with
  its precise meaning depending on the setting. In its most strict sense, it
  refers to a phase with spontaneous symmetry breaking, where a charged local
  operator acquires a nonzero expectation value. More broadly, it can also
  refer to the nonzero expectation values of scalar fields coupled to dynamical
  gauge fields and is used in the context of Higgs transitions. In addition, in
  ``anyon condensation,'' it denotes a formal mapping between topological
  quantum field theories. In this work, we use ``condensation'' in these
  standard, context-dependent senses; for further clarification and precise
  terminology, see, e.g., Ref. \protect \rev@citealp
  {cheng2026proliferation}.}\BibitemShut {Stop}%
\bibitem [{\citenamefont {Wegner}(1971)}]{Wegner}%
  \BibitemOpen
  \bibfield  {author} {\bibinfo {author} {\bibfnamefont {F.~J.}\ \bibnamefont
  {Wegner}},\ }\bibfield  {title} {\bibinfo {title} {Duality in generalized
  {Ising} models and phase transitions without local order parameters},\ }\href
  {https://doi.org/10.1063/1.1665530} {\bibfield  {journal} {\bibinfo
  {journal} {J. Math. Phys.}\ }\textbf {\bibinfo {volume} {12}},\ \bibinfo
  {pages} {2259} (\bibinfo {year} {1971})}\BibitemShut {NoStop}%
\bibitem [{\citenamefont {Fradkin}\ and\ \citenamefont
  {Shenker}(1979)}]{FradkinShenker}%
  \BibitemOpen
  \bibfield  {author} {\bibinfo {author} {\bibfnamefont {E.}~\bibnamefont
  {Fradkin}}\ and\ \bibinfo {author} {\bibfnamefont {S.~H.}\ \bibnamefont
  {Shenker}},\ }\bibfield  {title} {\bibinfo {title} {Phase diagrams of lattice
  gauge theories with {Higgs} fields},\ }\href
  {https://doi.org/10.1103/PhysRevD.19.3682} {\bibfield  {journal} {\bibinfo
  {journal} {Phys. Rev. D}\ }\textbf {\bibinfo {volume} {19}},\ \bibinfo
  {pages} {3682} (\bibinfo {year} {1979})}\BibitemShut {NoStop}%
\bibitem [{\citenamefont {{Tupitsyn}}\ \emph {et~al.}(2010)\citenamefont
  {{Tupitsyn}}, \citenamefont {{Kitaev}}, \citenamefont {{Prokof'Ev}},\ and\
  \citenamefont {{Stamp}}}]{TupitsynKitaevProkofEvStamp}%
  \BibitemOpen
  \bibfield  {author} {\bibinfo {author} {\bibfnamefont {I.~S.}\ \bibnamefont
  {{Tupitsyn}}}, \bibinfo {author} {\bibfnamefont {A.}~\bibnamefont
  {{Kitaev}}}, \bibinfo {author} {\bibfnamefont {N.~V.}\ \bibnamefont
  {{Prokof'Ev}}},\ and\ \bibinfo {author} {\bibfnamefont {P.~C.~E.}\
  \bibnamefont {{Stamp}}},\ }\bibfield  {title} {\bibinfo {title} {{Topological
  multicritical point in the phase diagram of the toric code model and
  three-dimensional lattice gauge Higgs model}},\ }\href
  {https://doi.org/10.1103/PhysRevB.82.085114} {\bibfield  {journal} {\bibinfo
  {journal} {\prb}\ }\textbf {\bibinfo {volume} {82}},\ \bibinfo {eid} {085114}
  (\bibinfo {year} {2010})},\ \Eprint {https://arxiv.org/abs/0804.3175}
  {arXiv:0804.3175 [cond-mat.stat-mech]} \BibitemShut {NoStop}%
\bibitem [{\citenamefont {{Vidal}}\ \emph {et~al.}(2009)\citenamefont
  {{Vidal}}, \citenamefont {{Dusuel}},\ and\ \citenamefont
  {{Schmidt}}}]{VidalDusuelSchmidt}%
  \BibitemOpen
  \bibfield  {author} {\bibinfo {author} {\bibfnamefont {J.}~\bibnamefont
  {{Vidal}}}, \bibinfo {author} {\bibfnamefont {S.}~\bibnamefont {{Dusuel}}},\
  and\ \bibinfo {author} {\bibfnamefont {K.~P.}\ \bibnamefont {{Schmidt}}},\
  }\bibfield  {title} {\bibinfo {title} {{Low-energy effective theory of the
  toric code model in a parallel magnetic field}},\ }\href
  {https://doi.org/10.1103/PhysRevB.79.033109} {\bibfield  {journal} {\bibinfo
  {journal} {\prb}\ }\textbf {\bibinfo {volume} {79}},\ \bibinfo {eid} {033109}
  (\bibinfo {year} {2009})},\ \Eprint {https://arxiv.org/abs/0807.0487}
  {arXiv:0807.0487 [cond-mat.other]} \BibitemShut {NoStop}%
\bibitem [{\citenamefont {{Wu}}\ \emph {et~al.}(2012)\citenamefont {{Wu}},
  \citenamefont {{Deng}},\ and\ \citenamefont {{Prokof'ev}}}]{WuDengProkofev}%
  \BibitemOpen
  \bibfield  {author} {\bibinfo {author} {\bibfnamefont {F.}~\bibnamefont
  {{Wu}}}, \bibinfo {author} {\bibfnamefont {Y.}~\bibnamefont {{Deng}}},\ and\
  \bibinfo {author} {\bibfnamefont {N.}~\bibnamefont {{Prokof'ev}}},\
  }\bibfield  {title} {\bibinfo {title} {{Phase diagram of the toric code model
  in a parallel magnetic field}},\ }\href
  {https://doi.org/10.1103/PhysRevB.85.195104} {\bibfield  {journal} {\bibinfo
  {journal} {\prb}\ }\textbf {\bibinfo {volume} {85}},\ \bibinfo {eid} {195104}
  (\bibinfo {year} {2012})},\ \Eprint {https://arxiv.org/abs/1201.6409}
  {arXiv:1201.6409 [cond-mat.stat-mech]} \BibitemShut {NoStop}%
\bibitem [{\citenamefont {{Somoza}}\ \emph {et~al.}(2021)\citenamefont
  {{Somoza}}, \citenamefont {{Serna}},\ and\ \citenamefont
  {{Nahum}}}]{SomozaSernaNahum}%
  \BibitemOpen
  \bibfield  {author} {\bibinfo {author} {\bibfnamefont {A.~M.}\ \bibnamefont
  {{Somoza}}}, \bibinfo {author} {\bibfnamefont {P.}~\bibnamefont {{Serna}}},\
  and\ \bibinfo {author} {\bibfnamefont {A.}~\bibnamefont {{Nahum}}},\
  }\bibfield  {title} {\bibinfo {title} {Self-dual criticality in
  three-dimensional $\mathbb{Z}_{2}$ gauge theory with matter},\ }\href
  {https://doi.org/10.1103/PhysRevX.11.041008} {\bibfield  {journal} {\bibinfo
  {journal} {Phys. Rev. X}\ }\textbf {\bibinfo {volume} {11}},\ \bibinfo {eid}
  {041008} (\bibinfo {year} {2021})},\ \Eprint
  {https://arxiv.org/abs/2012.15845} {arXiv:2012.15845 [cond-mat.stat-mech]}
  \BibitemShut {NoStop}%
\bibitem [{\citenamefont {{Oppenheim}}\ \emph {et~al.}(2024)\citenamefont
  {{Oppenheim}}, \citenamefont {{Koch-Janusz}}, \citenamefont {{Gazit}},\ and\
  \citenamefont {{Ringel}}}]{OppenheimKochJanuszGazitRingel}%
  \BibitemOpen
  \bibfield  {author} {\bibinfo {author} {\bibfnamefont {L.}~\bibnamefont
  {{Oppenheim}}}, \bibinfo {author} {\bibfnamefont {M.}~\bibnamefont
  {{Koch-Janusz}}}, \bibinfo {author} {\bibfnamefont {S.}~\bibnamefont
  {{Gazit}}},\ and\ \bibinfo {author} {\bibfnamefont {Z.}~\bibnamefont
  {{Ringel}}},\ }\bibfield  {title} {\bibinfo {title} {{Machine learning the
  operator content of the critical self-dual Ising-Higgs lattice gauge
  theory}},\ }\href {https://doi.org/10.1103/PhysRevResearch.6.043322}
  {\bibfield  {journal} {\bibinfo  {journal} {Phys. Rev. Research}\ }\textbf
  {\bibinfo {volume} {6}},\ \bibinfo {eid} {043322} (\bibinfo {year} {2024})},\
  \Eprint {https://arxiv.org/abs/2311.17994} {arXiv:2311.17994
  [cond-mat.str-el]} \BibitemShut {NoStop}%
\bibitem [{\citenamefont {{Xu}}\ \emph {et~al.}(2025)\citenamefont {{Xu}},
  \citenamefont {{Pollmann}},\ and\ \citenamefont {{Knap}}}]{XuPollmannKnap}%
  \BibitemOpen
  \bibfield  {author} {\bibinfo {author} {\bibfnamefont {W.-T.}\ \bibnamefont
  {{Xu}}}, \bibinfo {author} {\bibfnamefont {F.}~\bibnamefont {{Pollmann}}},\
  and\ \bibinfo {author} {\bibfnamefont {M.}~\bibnamefont {{Knap}}},\
  }\bibfield  {title} {\bibinfo {title} {{Critical behavior of
  Fredenhagen-Marcu string order parameters at topological phase transitions
  with emergent higher-form symmetries}},\ }\href
  {https://doi.org/10.1038/s41534-025-01030-z} {\bibfield  {journal} {\bibinfo
  {journal} {npj Quantum Information}\ }\textbf {\bibinfo {volume} {11}},\
  \bibinfo {eid} {74} (\bibinfo {year} {2025})},\ \Eprint
  {https://arxiv.org/abs/2402.00127} {arXiv:2402.00127 [cond-mat.str-el]}
  \BibitemShut {NoStop}%
\bibitem [{Note2()}]{Note2}%
  \BibitemOpen
  \bibinfo {note} {Actual numerical simulations often work with Euclidean
  lattice actions instead of Hamiltonians. The universal physics is expected to
  be the same.}\BibitemShut {Stop}%
\bibitem [{sup()}]{supp}%
  \BibitemOpen
  \href@noop {} {}\bibinfo {note} {See Supplemental Material at [URL] for a
  brief review of two existing proposals and the extent to which they meet both
  criteria discussed in the main text.}\BibitemShut {Stop}%
\bibitem [{\citenamefont {Kitaev}(2006)}]{Kitaev:2005hzj}%
  \BibitemOpen
  \bibfield  {author} {\bibinfo {author} {\bibfnamefont {A.}~\bibnamefont
  {Kitaev}},\ }\bibfield  {title} {\bibinfo {title} {{Anyons in an exactly
  solved model and beyond}},\ }\href
  {https://doi.org/10.1016/j.aop.2005.10.005} {\bibfield  {journal} {\bibinfo
  {journal} {Ann. Phys.}\ }\textbf {\bibinfo {volume} {321}},\ \bibinfo {pages}
  {2} (\bibinfo {year} {2006})},\ \Eprint
  {https://arxiv.org/abs/cond-mat/0506438} {arXiv:cond-mat/0506438}
  \BibitemShut {NoStop}%
\bibitem [{\citenamefont {{Bhardwaj}}\ \emph {et~al.}()\citenamefont
  {{Bhardwaj}}, \citenamefont {{Gaiotto}},\ and\ \citenamefont
  {{Kapustin}}}]{BhardwajGaiottoKapustin2017}%
  \BibitemOpen
  \bibfield  {author} {\bibinfo {author} {\bibfnamefont {L.}~\bibnamefont
  {{Bhardwaj}}}, \bibinfo {author} {\bibfnamefont {D.}~\bibnamefont
  {{Gaiotto}}},\ and\ \bibinfo {author} {\bibfnamefont {A.}~\bibnamefont
  {{Kapustin}}},\ }\bibfield  {title} {\bibinfo {title} {{State sum
  constructions of spin-TFTs and string net constructions of fermionic phases
  of matter}},\ }\href {https://doi.org/10.1007/JHEP04(2017)096} {\bibfield
  {journal} {\bibinfo  {journal} {J. High Energy Phys.}\ }\textbf {\bibinfo
  {volume} {04}}\bibfield  {number} {\bibinfo  {number} { (2017)},\ \bibinfo
  {pages} {096}},\ }\Eprint {https://arxiv.org/abs/1605.01640}
  {arXiv:1605.01640 [cond-mat.str-el]} \BibitemShut {NoStop}%
\bibitem [{\citenamefont {{Bais}}\ and\ \citenamefont
  {{Slingerland}}(2009)}]{bais2009condensate}%
  \BibitemOpen
  \bibfield  {author} {\bibinfo {author} {\bibfnamefont {F.~A.}\ \bibnamefont
  {{Bais}}}\ and\ \bibinfo {author} {\bibfnamefont {J.~K.}\ \bibnamefont
  {{Slingerland}}},\ }\bibfield  {title} {\bibinfo {title} {{Condensate-induced
  transitions between topologically ordered phases}},\ }\href
  {https://doi.org/10.1103/PhysRevB.79.045316} {\bibfield  {journal} {\bibinfo
  {journal} {\prb}\ }\textbf {\bibinfo {volume} {79}},\ \bibinfo {eid} {045316}
  (\bibinfo {year} {2009})},\ \Eprint {https://arxiv.org/abs/0808.0627}
  {arXiv:0808.0627 [cond-mat.mes-hall]} \BibitemShut {NoStop}%
\bibitem [{\citenamefont {Teo}\ \emph {et~al.}(2015)\citenamefont {Teo},
  \citenamefont {Hughes},\ and\ \citenamefont {Fradkin}}]{TeoHughesFradkin}%
  \BibitemOpen
  \bibfield  {author} {\bibinfo {author} {\bibfnamefont {J.~C.}\ \bibnamefont
  {Teo}}, \bibinfo {author} {\bibfnamefont {T.~L.}\ \bibnamefont {Hughes}},\
  and\ \bibinfo {author} {\bibfnamefont {E.}~\bibnamefont {Fradkin}},\
  }\bibfield  {title} {\bibinfo {title} {Theory of twist liquids: Gauging an
  anyonic symmetry},\ }\href
  {https://doi.org/https://doi.org/10.1016/j.aop.2015.05.012} {\bibfield
  {journal} {\bibinfo  {journal} {Ann. Phys.}\ }\textbf {\bibinfo {volume}
  {360}},\ \bibinfo {pages} {349} (\bibinfo {year} {2015})},\ \Eprint
  {https://arxiv.org/abs/1503.06812} {arXiv:1503.06812 [cond-mat.str-el]}
  \BibitemShut {NoStop}%
\bibitem [{Note3()}]{Note3}%
  \BibitemOpen
  \bibinfo {note} {Naively there are more: $\nu $ can take any odd integer mod
  $16$ \cite {Kitaev:2005hzj}. Obviously $\nu $ is equivalent to $-\nu $ since
  the theory is doubled. It turns out that $\nu $ is also equivalent to $\nu
  +8$ through the following anyon relabeling: $\sigma '=\sigma \times \protect
  \bar {\psi }$ and $\protect \bar {\sigma }'=\protect \bar {\sigma }\times
  \psi $. Then we are left with only two distinct possibilities.}\BibitemShut
  {Stop}%
\bibitem [{\citenamefont {Levin}\ and\ \citenamefont {Gu}(2012)}]{LevinGu}%
  \BibitemOpen
  \bibfield  {author} {\bibinfo {author} {\bibfnamefont {M.}~\bibnamefont
  {Levin}}\ and\ \bibinfo {author} {\bibfnamefont {Z.-C.}\ \bibnamefont {Gu}},\
  }\bibfield  {title} {\bibinfo {title} {Braiding statistics approach to
  symmetry-protected topological phases},\ }\href
  {https://journals.aps.org/prb/abstract/10.1103/PhysRevB.86.115109} {\bibfield
   {journal} {\bibinfo  {journal} {Phys. Rev. B}\ }\textbf {\bibinfo {volume}
  {86}},\ \bibinfo {pages} {115109} (\bibinfo {year} {2012})},\ \Eprint
  {https://arxiv.org/abs/1202.3120} {arXiv:1202.3120 [cond-mat.str-el]}
  \BibitemShut {NoStop}%
\bibitem [{\citenamefont {{Gaiotto}}\ \emph {et~al.}()\citenamefont
  {{Gaiotto}}, \citenamefont {{Kapustin}}, \citenamefont {{Seiberg}},\ and\
  \citenamefont {{Willett}}}]{HigherFormSym}%
  \BibitemOpen
  \bibfield  {author} {\bibinfo {author} {\bibfnamefont {D.}~\bibnamefont
  {{Gaiotto}}}, \bibinfo {author} {\bibfnamefont {A.}~\bibnamefont
  {{Kapustin}}}, \bibinfo {author} {\bibfnamefont {N.}~\bibnamefont
  {{Seiberg}}},\ and\ \bibinfo {author} {\bibfnamefont {B.}~\bibnamefont
  {{Willett}}},\ }\bibfield  {title} {\bibinfo {title} {{Generalized global
  symmetries}},\ }\href {https://doi.org/10.1007/JHEP02(2015)172} {\bibfield
  {journal} {\bibinfo  {journal} {J. High Energy Phys.}\ }\textbf {\bibinfo
  {volume} {02}}\bibfield  {number} {\bibinfo  {number} { (2015)},\ \bibinfo
  {pages} {172}},\ }\Eprint {https://arxiv.org/abs/1412.5148} {arXiv:1412.5148
  [hep-th]} \BibitemShut {NoStop}%
\bibitem [{\citenamefont {{Kovner}}(2002)}]{KovnerReview}%
  \BibitemOpen
  \bibfield  {author} {\bibinfo {author} {\bibfnamefont {A.}~\bibnamefont
  {{Kovner}}},\ }\bibfield  {title} {\bibinfo {title} {Magnetic ${Z}_{N}$
  symmetry in $2+1$ dimensions},\ }\href
  {https://doi.org/10.1142/S0217751X02010789} {\bibfield  {journal} {\bibinfo
  {journal} {Int. J. Mod. Phys. A}\ }\textbf {\bibinfo {volume} {17}},\
  \bibinfo {pages} {2113} (\bibinfo {year} {2002})},\ \Eprint
  {https://arxiv.org/abs/hep-th/0211248} {arXiv:hep-th/0211248 [hep-th]}
  \BibitemShut {NoStop}%
\bibitem [{\citenamefont {{Aharony}}\ \emph {et~al.}()\citenamefont
  {{Aharony}}, \citenamefont {{Seiberg}},\ and\ \citenamefont
  {{Tachikawa}}}]{ReadingBetweenLines}%
  \BibitemOpen
  \bibfield  {author} {\bibinfo {author} {\bibfnamefont {O.}~\bibnamefont
  {{Aharony}}}, \bibinfo {author} {\bibfnamefont {N.}~\bibnamefont
  {{Seiberg}}},\ and\ \bibinfo {author} {\bibfnamefont {Y.}~\bibnamefont
  {{Tachikawa}}},\ }\bibfield  {title} {\bibinfo {title} {{Reading between the
  lines of four-dimensional gauge theories}},\ }\href
  {https://doi.org/10.1007/JHEP08(2013)115} {\bibfield  {journal} {\bibinfo
  {journal} {J. High Energy Phys.}\ }\textbf {\bibinfo {volume} {08}}\bibfield
  {number} {\bibinfo  {number} { (2013)},\ \bibinfo {pages} {115}},\ }\Eprint
  {https://arxiv.org/abs/1305.0318} {arXiv:1305.0318 [hep-th]} \BibitemShut
  {NoStop}%
\bibitem [{\citenamefont {{Teper}}(1998)}]{Teper}%
  \BibitemOpen
  \bibfield  {author} {\bibinfo {author} {\bibfnamefont {M.~J.}\ \bibnamefont
  {{Teper}}},\ }\bibfield  {title} {\bibinfo {title} {${SU(N)}$ gauge theories
  in 2+1 dimensions},\ }\href {https://doi.org/10.1103/PhysRevD.59.014512}
  {\bibfield  {journal} {\bibinfo  {journal} {\prd}\ }\textbf {\bibinfo
  {volume} {59}},\ \bibinfo {eid} {014512} (\bibinfo {year} {1998})},\ \Eprint
  {https://arxiv.org/abs/hep-lat/9804008} {arXiv:hep-lat/9804008 [hep-lat]}
  \BibitemShut {NoStop}%
\bibitem [{\citenamefont {{Athenodorou}}\ and\ \citenamefont
  {{Teper}}(2017)}]{AthenodorouTeper}%
  \BibitemOpen
  \bibfield  {author} {\bibinfo {author} {\bibfnamefont {A.}~\bibnamefont
  {{Athenodorou}}}\ and\ \bibinfo {author} {\bibfnamefont {M.}~\bibnamefont
  {{Teper}}},\ }\bibfield  {title} {\bibinfo {title} {{$SU(N)$} gauge theories
  in 2+1 dimensions: glueball spectra and $k$-string tensions},\ }\href
  {https://doi.org/10.1007/JHEP02(2017)015} {\bibfield  {journal} {\bibinfo
  {journal} {J. High Energy Phys.}\ }\textbf {\bibinfo {volume} {02}}\bibfield
  {number} {\bibinfo  {number} { (2017)},\ \bibinfo {pages} {015}},\ }\Eprint
  {https://arxiv.org/abs/1609.03873} {arXiv:1609.03873 [hep-lat]} \BibitemShut
  {NoStop}%
\bibitem [{\citenamefont {Dasgupta}\ and\ \citenamefont
  {Halperin}(1981)}]{DasguptaHalperin}%
  \BibitemOpen
  \bibfield  {author} {\bibinfo {author} {\bibfnamefont {C.}~\bibnamefont
  {Dasgupta}}\ and\ \bibinfo {author} {\bibfnamefont {B.~I.}\ \bibnamefont
  {Halperin}},\ }\bibfield  {title} {\bibinfo {title} {Phase transition in a
  lattice model of superconductivity},\ }\href
  {https://doi.org/10.1103/PhysRevLett.47.1556} {\bibfield  {journal} {\bibinfo
   {journal} {Phys. Rev. Lett.}\ }\textbf {\bibinfo {volume} {47}},\ \bibinfo
  {pages} {1556} (\bibinfo {year} {1981})}\BibitemShut {NoStop}%
\bibitem [{\citenamefont {Coleman}\ and\ \citenamefont
  {Weinberg}(1973)}]{Coleman1973}%
  \BibitemOpen
  \bibfield  {author} {\bibinfo {author} {\bibfnamefont {S.~R.}\ \bibnamefont
  {Coleman}}\ and\ \bibinfo {author} {\bibfnamefont {E.~J.}\ \bibnamefont
  {Weinberg}},\ }\bibfield  {title} {\bibinfo {title} {Radiative corrections as
  the origin of spontaneous symmetry breaking},\ }\href
  {https://doi.org/10.1103/PhysRevD.7.1888} {\bibfield  {journal} {\bibinfo
  {journal} {Phys. Rev. D}\ }\textbf {\bibinfo {volume} {7}},\ \bibinfo {pages}
  {1888} (\bibinfo {year} {1973})}\BibitemShut {NoStop}%
\bibitem [{\citenamefont {Halperin}\ \emph {et~al.}(1974)\citenamefont
  {Halperin}, \citenamefont {Lubensky},\ and\ \citenamefont
  {Ma}}]{Halperin1974}%
  \BibitemOpen
  \bibfield  {author} {\bibinfo {author} {\bibfnamefont {B.~I.}\ \bibnamefont
  {Halperin}}, \bibinfo {author} {\bibfnamefont {T.~C.}\ \bibnamefont
  {Lubensky}},\ and\ \bibinfo {author} {\bibfnamefont {S.-k.}\ \bibnamefont
  {Ma}},\ }\bibfield  {title} {\bibinfo {title} {First-order phase transitions
  in superconductors and smectic-a liquid crystals},\ }\href
  {https://doi.org/10.1103/PhysRevLett.32.292} {\bibfield  {journal} {\bibinfo
  {journal} {Phys. Rev. Lett.}\ }\textbf {\bibinfo {volume} {32}},\ \bibinfo
  {pages} {292} (\bibinfo {year} {1974})}\BibitemShut {NoStop}%
\bibitem [{\citenamefont {{Kos}}\ \emph {et~al.}()\citenamefont {{Kos}},
  \citenamefont {{Poland}}, \citenamefont {{Simmons-Duffin}},\ and\
  \citenamefont {{Vichi}}}]{O(N)Bootstrap}%
  \BibitemOpen
  \bibfield  {author} {\bibinfo {author} {\bibfnamefont {F.}~\bibnamefont
  {{Kos}}}, \bibinfo {author} {\bibfnamefont {D.}~\bibnamefont {{Poland}}},
  \bibinfo {author} {\bibfnamefont {D.}~\bibnamefont {{Simmons-Duffin}}},\ and\
  \bibinfo {author} {\bibfnamefont {A.}~\bibnamefont {{Vichi}}},\ }\bibfield
  {title} {\bibinfo {title} {{Bootstrapping the ${O}({N})$ archipelago}},\
  }\href {https://doi.org/10.1007/JHEP11(2015)106} {\bibfield  {journal}
  {\bibinfo  {journal} {J. High Energy Phys.}\ }\textbf {\bibinfo {volume}
  {11}}\bibfield  {number} {\bibinfo  {number} { (2015)},\ \bibinfo {pages}
  {106}},\ }\Eprint {https://arxiv.org/abs/1504.07997} {arXiv:1504.07997
  [hep-th]} \BibitemShut {NoStop}%
\bibitem [{\citenamefont {{Hasenbusch}}(2001)}]{O(4)MonteCarlo}%
  \BibitemOpen
  \bibfield  {author} {\bibinfo {author} {\bibfnamefont {M.}~\bibnamefont
  {{Hasenbusch}}},\ }\bibfield  {title} {\bibinfo {title} {Eliminating leading
  corrections to scaling in the three-dimensional ${O}({N})$-symmetric
  {\ensuremath{\varphi}}$^{4}$ model: ${N} = 3$ and $4$},\ }\href
  {https://doi.org/10.1088/0305-4470/34/40/302} {\bibfield  {journal} {\bibinfo
   {journal} {J. Phys. A}\ }\textbf {\bibinfo {volume} {34}},\ \bibinfo {pages}
  {8221} (\bibinfo {year} {2001})},\ \Eprint
  {https://arxiv.org/abs/cond-mat/0010463} {arXiv:cond-mat/0010463 [cond-mat]}
  \BibitemShut {NoStop}%
\bibitem [{Note4()}]{Note4}%
  \BibitemOpen
  \bibinfo {note} {More accurately, one must additionally tune the mass term
  $\Phi ^2$ to reach the Ising$^*$ critical point. The shape of the transition
  line is then controlled by the scaling dimensions of the mass and monopole
  operators.}\BibitemShut {Stop}%
\bibitem [{\citenamefont {Senthil}\ \emph
  {et~al.}(2004{\natexlab{a}})\citenamefont {Senthil}, \citenamefont
  {Vishwanath}, \citenamefont {Balents}, \citenamefont {Sachdev},\ and\
  \citenamefont {Fisher}}]{deconfine1}%
  \BibitemOpen
  \bibfield  {author} {\bibinfo {author} {\bibfnamefont {T.}~\bibnamefont
  {Senthil}}, \bibinfo {author} {\bibfnamefont {A.}~\bibnamefont {Vishwanath}},
  \bibinfo {author} {\bibfnamefont {L.}~\bibnamefont {Balents}}, \bibinfo
  {author} {\bibfnamefont {S.}~\bibnamefont {Sachdev}},\ and\ \bibinfo {author}
  {\bibfnamefont {M.~P.~A.}\ \bibnamefont {Fisher}},\ }\bibfield  {title}
  {\bibinfo {title} {Deconfined quantum critical points},\ }\href
  {https://doi.org/10.1126/science.1091806} {\bibfield  {journal} {\bibinfo
  {journal} {Science}\ }\textbf {\bibinfo {volume} {303}},\ \bibinfo {pages}
  {1490} (\bibinfo {year} {2004}{\natexlab{a}})},\ \Eprint
  {https://arxiv.org/abs/cond-mat/0311326} {arXiv:cond-mat/0311326}
  \BibitemShut {NoStop}%
\bibitem [{\citenamefont {Senthil}\ \emph
  {et~al.}(2004{\natexlab{b}})\citenamefont {Senthil}, \citenamefont {Balents},
  \citenamefont {Sachdev}, \citenamefont {Vishwanath},\ and\ \citenamefont
  {Fisher}}]{deconfine2}%
  \BibitemOpen
  \bibfield  {author} {\bibinfo {author} {\bibfnamefont {T.}~\bibnamefont
  {Senthil}}, \bibinfo {author} {\bibfnamefont {L.}~\bibnamefont {Balents}},
  \bibinfo {author} {\bibfnamefont {S.}~\bibnamefont {Sachdev}}, \bibinfo
  {author} {\bibfnamefont {A.}~\bibnamefont {Vishwanath}},\ and\ \bibinfo
  {author} {\bibfnamefont {M.~P.~A.}\ \bibnamefont {Fisher}},\ }\bibfield
  {title} {\bibinfo {title} {Quantum criticality beyond the
  {Landau-Ginzburg-Wilson} paradigm},\ }\href
  {https://doi.org/10.1103/PhysRevB.70.144407} {\bibfield  {journal} {\bibinfo
  {journal} {Phys. Rev. B}\ }\textbf {\bibinfo {volume} {70}},\ \bibinfo
  {pages} {144407} (\bibinfo {year} {2004}{\natexlab{b}})}\BibitemShut
  {NoStop}%
\bibitem [{\citenamefont {Beigi}\ \emph {et~al.}(2011)\citenamefont {Beigi},
  \citenamefont {Shor},\ and\ \citenamefont {Whalen}}]{beigi2011quantum}%
  \BibitemOpen
  \bibfield  {author} {\bibinfo {author} {\bibfnamefont {S.}~\bibnamefont
  {Beigi}}, \bibinfo {author} {\bibfnamefont {P.~W.}\ \bibnamefont {Shor}},\
  and\ \bibinfo {author} {\bibfnamefont {D.}~\bibnamefont {Whalen}},\
  }\bibfield  {title} {\bibinfo {title} {{The Quantum Double Model with
  Boundary: Condensations and Symmetries}},\ }\href
  {https://doi.org/10.1007/s00220-011-1294-x} {\bibfield  {journal} {\bibinfo
  {journal} {Commun. Math. Phys.}\ }\textbf {\bibinfo {volume} {306}},\
  \bibinfo {pages} {663} (\bibinfo {year} {2011})},\ \Eprint
  {https://arxiv.org/abs/1006.5479} {arXiv:1006.5479 [quant-ph]} \BibitemShut
  {NoStop}%
\bibitem [{\citenamefont {{Senthil}}\ \emph {et~al.}(2019)\citenamefont
  {{Senthil}}, \citenamefont {{Son}}, \citenamefont {{Wang}},\ and\
  \citenamefont {{Xu}}}]{DualityReview}%
  \BibitemOpen
  \bibfield  {author} {\bibinfo {author} {\bibfnamefont {T.}~\bibnamefont
  {{Senthil}}}, \bibinfo {author} {\bibfnamefont {D.~T.}\ \bibnamefont
  {{Son}}}, \bibinfo {author} {\bibfnamefont {C.}~\bibnamefont {{Wang}}},\ and\
  \bibinfo {author} {\bibfnamefont {C.}~\bibnamefont {{Xu}}},\ }\bibfield
  {title} {\bibinfo {title} {{Duality between $(2 + 1) d$ quantum critical
  points}},\ }\href {https://doi.org/10.1016/j.physrep.2019.09.001} {\bibfield
  {journal} {\bibinfo  {journal} {Phys. Rep.}\ }\textbf {\bibinfo {volume}
  {827}},\ \bibinfo {pages} {1} (\bibinfo {year} {2019})},\ \Eprint
  {https://arxiv.org/abs/1810.05174} {arXiv:1810.05174 [cond-mat.str-el]}
  \BibitemShut {NoStop}%
\bibitem [{\citenamefont {{Nahum}}\ \emph {et~al.}(2015)\citenamefont
  {{Nahum}}, \citenamefont {{Serna}}, \citenamefont {{Chalker}}, \citenamefont
  {{Ortu{\~n}o}},\ and\ \citenamefont {{Somoza}}}]{SO(5)}%
  \BibitemOpen
  \bibfield  {author} {\bibinfo {author} {\bibfnamefont {A.}~\bibnamefont
  {{Nahum}}}, \bibinfo {author} {\bibfnamefont {P.}~\bibnamefont {{Serna}}},
  \bibinfo {author} {\bibfnamefont {J.~T.}\ \bibnamefont {{Chalker}}}, \bibinfo
  {author} {\bibfnamefont {M.}~\bibnamefont {{Ortu{\~n}o}}},\ and\ \bibinfo
  {author} {\bibfnamefont {A.~M.}\ \bibnamefont {{Somoza}}},\ }\bibfield
  {title} {\bibinfo {title} {Emergent $\textrm{SO}(5)$ symmetry at the
  {N{\'e}el} to valence-bond-solid transition},\ }\href
  {https://doi.org/10.1103/PhysRevLett.115.267203} {\bibfield  {journal}
  {\bibinfo  {journal} {\prl}\ }\textbf {\bibinfo {volume} {115}},\ \bibinfo
  {eid} {267203} (\bibinfo {year} {2015})},\ \Eprint
  {https://arxiv.org/abs/1508.06668} {arXiv:1508.06668 [cond-mat.str-el]}
  \BibitemShut {NoStop}%
\bibitem [{\citenamefont {Wang}\ \emph {et~al.}(2017)\citenamefont {Wang},
  \citenamefont {Nahum}, \citenamefont {Metlitski}, \citenamefont {Xu},\ and\
  \citenamefont {Senthil}}]{SO5}%
  \BibitemOpen
  \bibfield  {author} {\bibinfo {author} {\bibfnamefont {C.}~\bibnamefont
  {Wang}}, \bibinfo {author} {\bibfnamefont {A.}~\bibnamefont {Nahum}},
  \bibinfo {author} {\bibfnamefont {M.~A.}\ \bibnamefont {Metlitski}}, \bibinfo
  {author} {\bibfnamefont {C.}~\bibnamefont {Xu}},\ and\ \bibinfo {author}
  {\bibfnamefont {T.}~\bibnamefont {Senthil}},\ }\bibfield  {title} {\bibinfo
  {title} {Deconfined quantum critical points: Symmetries and dualities},\
  }\href {https://doi.org/10.1103/PhysRevX.7.031051} {\bibfield  {journal}
  {\bibinfo  {journal} {Phys. Rev. X}\ }\textbf {\bibinfo {volume} {7}},\
  \bibinfo {pages} {031051} (\bibinfo {year} {2017})},\ \Eprint
  {https://arxiv.org/abs/1703.02426} {arXiv:1703.02426 [cond-mat.str-el]}
  \BibitemShut {NoStop}%
\bibitem [{\citenamefont {Kos}\ \emph {et~al.}(2014)\citenamefont {Kos},
  \citenamefont {Poland},\ and\ \citenamefont {Simmons-Duffin}}]{Kos:2014bka}%
  \BibitemOpen
  \bibfield  {author} {\bibinfo {author} {\bibfnamefont {F.}~\bibnamefont
  {Kos}}, \bibinfo {author} {\bibfnamefont {D.}~\bibnamefont {Poland}},\ and\
  \bibinfo {author} {\bibfnamefont {D.}~\bibnamefont {Simmons-Duffin}},\
  }\bibfield  {title} {\bibinfo {title} {{Bootstrapping Mixed Correlators in
  the 3D Ising Model}},\ }\href {https://doi.org/10.1007/JHEP11(2014)109}
  {\bibfield  {journal} {\bibinfo  {journal} {JHEP}\ }\textbf {\bibinfo
  {volume} {11}},\ \bibinfo {pages} {109}},\ \Eprint
  {https://arxiv.org/abs/1406.4858} {arXiv:1406.4858 [hep-th]} \BibitemShut
  {NoStop}%
\bibitem [{\citenamefont {El-Showk}\ \emph {et~al.}(2012)\citenamefont
  {El-Showk}, \citenamefont {Paulos}, \citenamefont {Poland}, \citenamefont
  {Rychkov}, \citenamefont {Simmons-Duffin},\ and\ \citenamefont
  {Vichi}}]{El-Showk:2012cjh}%
  \BibitemOpen
  \bibfield  {author} {\bibinfo {author} {\bibfnamefont {S.}~\bibnamefont
  {El-Showk}}, \bibinfo {author} {\bibfnamefont {M.~F.}\ \bibnamefont
  {Paulos}}, \bibinfo {author} {\bibfnamefont {D.}~\bibnamefont {Poland}},
  \bibinfo {author} {\bibfnamefont {S.}~\bibnamefont {Rychkov}}, \bibinfo
  {author} {\bibfnamefont {D.}~\bibnamefont {Simmons-Duffin}},\ and\ \bibinfo
  {author} {\bibfnamefont {A.}~\bibnamefont {Vichi}},\ }\bibfield  {title}
  {\bibinfo {title} {{Solving the 3D Ising Model with the Conformal
  Bootstrap}},\ }\href {https://doi.org/10.1103/PhysRevD.86.025022} {\bibfield
  {journal} {\bibinfo  {journal} {Phys. Rev. D}\ }\textbf {\bibinfo {volume}
  {86}},\ \bibinfo {pages} {025022} (\bibinfo {year} {2012})},\ \Eprint
  {https://arxiv.org/abs/1203.6064} {arXiv:1203.6064 [hep-th]} \BibitemShut
  {NoStop}%
\bibitem [{\citenamefont {Chang}\ \emph {et~al.}(2025)\citenamefont {Chang},
  \citenamefont {Dommes}, \citenamefont {Erramilli}, \citenamefont {Homrich},
  \citenamefont {Kravchuk}, \citenamefont {Liu}, \citenamefont {Mitchell},
  \citenamefont {Poland},\ and\ \citenamefont
  {Simmons-Duffin}}]{Chang:2024whx}%
  \BibitemOpen
  \bibfield  {author} {\bibinfo {author} {\bibfnamefont {C.-H.}\ \bibnamefont
  {Chang}}, \bibinfo {author} {\bibfnamefont {V.}~\bibnamefont {Dommes}},
  \bibinfo {author} {\bibfnamefont {R.~S.}\ \bibnamefont {Erramilli}}, \bibinfo
  {author} {\bibfnamefont {A.}~\bibnamefont {Homrich}}, \bibinfo {author}
  {\bibfnamefont {P.}~\bibnamefont {Kravchuk}}, \bibinfo {author}
  {\bibfnamefont {A.}~\bibnamefont {Liu}}, \bibinfo {author} {\bibfnamefont
  {M.~S.}\ \bibnamefont {Mitchell}}, \bibinfo {author} {\bibfnamefont
  {D.}~\bibnamefont {Poland}},\ and\ \bibinfo {author} {\bibfnamefont
  {D.}~\bibnamefont {Simmons-Duffin}},\ }\bibfield  {title} {\bibinfo {title}
  {{Bootstrapping the 3d Ising stress tensor}},\ }\href
  {https://doi.org/10.1007/JHEP03(2025)136} {\bibfield  {journal} {\bibinfo
  {journal} {JHEP}\ }\textbf {\bibinfo {volume} {03}},\ \bibinfo {pages}
  {136}},\ \Eprint {https://arxiv.org/abs/2411.15300} {arXiv:2411.15300
  [hep-th]} \BibitemShut {NoStop}%
\bibitem [{\citenamefont {Allais}(2014)}]{allais2014magnetic}%
  \BibitemOpen
  \bibfield  {author} {\bibinfo {author} {\bibfnamefont {A.}~\bibnamefont
  {Allais}},\ }\href@noop {} {\bibinfo {title} {Magnetic defect line in a
  critical ising bath}} (\bibinfo {year} {2014}),\ \Eprint
  {https://arxiv.org/abs/1412.3449} {arXiv:1412.3449 [cond-mat.str-el]}
  \BibitemShut {NoStop}%
\bibitem [{\citenamefont {Allais}\ and\ \citenamefont
  {Sachdev}(2014)}]{Allais:2014fqa}%
  \BibitemOpen
  \bibfield  {author} {\bibinfo {author} {\bibfnamefont {A.}~\bibnamefont
  {Allais}}\ and\ \bibinfo {author} {\bibfnamefont {S.}~\bibnamefont
  {Sachdev}},\ }\bibfield  {title} {\bibinfo {title} {{Spectral function of a
  localized fermion coupled to the Wilson-Fisher conformal field theory}},\
  }\href {https://doi.org/10.1103/PhysRevB.90.035131} {\bibfield  {journal}
  {\bibinfo  {journal} {Phys. Rev. B}\ }\textbf {\bibinfo {volume} {90}},\
  \bibinfo {pages} {035131} (\bibinfo {year} {2014})},\ \Eprint
  {https://arxiv.org/abs/1406.3022} {arXiv:1406.3022 [cond-mat.str-el]}
  \BibitemShut {NoStop}%
\bibitem [{\citenamefont {Parisen~Toldin}\ \emph {et~al.}(2017)\citenamefont
  {Parisen~Toldin}, \citenamefont {Assaad},\ and\ \citenamefont
  {Wessel}}]{Parisen_Toldin_2017}%
  \BibitemOpen
  \bibfield  {author} {\bibinfo {author} {\bibfnamefont {F.}~\bibnamefont
  {Parisen~Toldin}}, \bibinfo {author} {\bibfnamefont {F.~F.}\ \bibnamefont
  {Assaad}},\ and\ \bibinfo {author} {\bibfnamefont {S.}~\bibnamefont
  {Wessel}},\ }\bibfield  {title} {\bibinfo {title} {Critical behavior in the
  presence of an order-parameter pinning field},\ }\bibfield  {journal}
  {\bibinfo  {journal} {Physical Review B}\ }\textbf {\bibinfo {volume} {95}},\
  \href {https://doi.org/10.1103/physrevb.95.014401}
  {10.1103/physrevb.95.014401} (\bibinfo {year} {2017})\BibitemShut {NoStop}%
\bibitem [{\citenamefont {Cuomo}\ \emph {et~al.}(2022)\citenamefont {Cuomo},
  \citenamefont {Komargodski},\ and\ \citenamefont {Mezei}}]{Cuomo:2021kfm}%
  \BibitemOpen
  \bibfield  {author} {\bibinfo {author} {\bibfnamefont {G.}~\bibnamefont
  {Cuomo}}, \bibinfo {author} {\bibfnamefont {Z.}~\bibnamefont {Komargodski}},\
  and\ \bibinfo {author} {\bibfnamefont {M.}~\bibnamefont {Mezei}},\ }\bibfield
   {title} {\bibinfo {title} {{Localized magnetic field in the O(N) model}},\
  }\href {https://doi.org/10.1007/JHEP02(2022)134} {\bibfield  {journal}
  {\bibinfo  {journal} {JHEP}\ }\textbf {\bibinfo {volume} {02}}\bibfield
  {number} {\bibinfo  {number} { (134)}},\ }\Eprint
  {https://arxiv.org/abs/2112.10634} {arXiv:2112.10634 [hep-th]} \BibitemShut
  {NoStop}%
\bibitem [{\citenamefont {Gimenez-Grau}(2022)}]{Gimenez-Grau:2022ebb}%
  \BibitemOpen
  \bibfield  {author} {\bibinfo {author} {\bibfnamefont {A.}~\bibnamefont
  {Gimenez-Grau}},\ }\href@noop {} {\bibinfo {title} {{Probing magnetic line
  defects with two-point functions}}} (\bibinfo {year} {2022}),\ \Eprint
  {https://arxiv.org/abs/2212.02520} {arXiv:2212.02520 [hep-th]} \BibitemShut
  {NoStop}%
\bibitem [{\citenamefont {{Bianchi}}\ \emph {et~al.}(2023)\citenamefont
  {{Bianchi}}, \citenamefont {{Bonomi}},\ and\ \citenamefont {{de
  Sabbata}}}]{BianchiBootstrapMagneticField}%
  \BibitemOpen
  \bibfield  {author} {\bibinfo {author} {\bibfnamefont {L.}~\bibnamefont
  {{Bianchi}}}, \bibinfo {author} {\bibfnamefont {D.}~\bibnamefont
  {{Bonomi}}},\ and\ \bibinfo {author} {\bibfnamefont {E.}~\bibnamefont {{de
  Sabbata}}},\ }\bibfield  {title} {\bibinfo {title} {{Analytic bootstrap for
  the localized magnetic field}},\ }\href
  {https://doi.org/10.1007/JHEP04(2023)069} {\bibfield  {journal} {\bibinfo
  {journal} {Journal of High Energy Physics}\ }\textbf {\bibinfo {volume}
  {2023}},\ \bibinfo {eid} {69} (\bibinfo {year} {2023})},\ \Eprint
  {https://arxiv.org/abs/2212.02524} {arXiv:2212.02524 [hep-th]} \BibitemShut
  {NoStop}%
\bibitem [{\citenamefont {Nishioka}\ \emph {et~al.}(2023)\citenamefont
  {Nishioka}, \citenamefont {Okuyama},\ and\ \citenamefont
  {Shimamori}}]{Nishioka:2022qmj}%
  \BibitemOpen
  \bibfield  {author} {\bibinfo {author} {\bibfnamefont {T.}~\bibnamefont
  {Nishioka}}, \bibinfo {author} {\bibfnamefont {Y.}~\bibnamefont {Okuyama}},\
  and\ \bibinfo {author} {\bibfnamefont {S.}~\bibnamefont {Shimamori}},\
  }\bibfield  {title} {\bibinfo {title} {{The epsilon expansion of the O(N)
  model with line defect from conformal field theory}},\ }\href
  {https://doi.org/10.1007/JHEP03(2023)203} {\bibfield  {journal} {\bibinfo
  {journal} {JHEP}\ }\textbf {\bibinfo {volume} {03}}\bibfield  {number}
  {\bibinfo  {number} { (203)}},\ }\Eprint {https://arxiv.org/abs/2212.04076}
  {arXiv:2212.04076 [hep-th]} \BibitemShut {NoStop}%
\bibitem [{\citenamefont {Hu}\ \emph {et~al.}(2024)\citenamefont {Hu},
  \citenamefont {He},\ and\ \citenamefont {Zhu}}]{Hu_2024}%
  \BibitemOpen
  \bibfield  {author} {\bibinfo {author} {\bibfnamefont {L.}~\bibnamefont
  {Hu}}, \bibinfo {author} {\bibfnamefont {Y.-C.}\ \bibnamefont {He}},\ and\
  \bibinfo {author} {\bibfnamefont {W.}~\bibnamefont {Zhu}},\ }\bibfield
  {title} {\bibinfo {title} {Solving conformal defects in 3d conformal field
  theory using fuzzy sphere regularization},\ }\bibfield  {journal} {\bibinfo
  {journal} {Nature Communications}\ }\textbf {\bibinfo {volume} {15}},\ \href
  {https://doi.org/10.1038/s41467-024-47978-y} {10.1038/s41467-024-47978-y}
  (\bibinfo {year} {2024})\BibitemShut {NoStop}%
\bibitem [{\citenamefont {Zhou}\ \emph {et~al.}(2024)\citenamefont {Zhou},
  \citenamefont {Gaiotto}, \citenamefont {He},\ and\ \citenamefont
  {Zou}}]{Zhou:2023fqu}%
  \BibitemOpen
  \bibfield  {author} {\bibinfo {author} {\bibfnamefont {Z.}~\bibnamefont
  {Zhou}}, \bibinfo {author} {\bibfnamefont {D.}~\bibnamefont {Gaiotto}},
  \bibinfo {author} {\bibfnamefont {Y.-C.}\ \bibnamefont {He}},\ and\ \bibinfo
  {author} {\bibfnamefont {Y.}~\bibnamefont {Zou}},\ }\bibfield  {title}
  {\bibinfo {title} {{The $g$-function and defect changing operators from
  wavefunction overlap on a fuzzy sphere}},\ }\href
  {https://doi.org/10.21468/SciPostPhys.17.1.021} {\bibfield  {journal}
  {\bibinfo  {journal} {SciPost Phys.}\ }\textbf {\bibinfo {volume} {17}},\
  \bibinfo {pages} {021} (\bibinfo {year} {2024})},\ \Eprint
  {https://arxiv.org/abs/2401.00039} {arXiv:2401.00039 [hep-th]} \BibitemShut
  {NoStop}%
\bibitem [{\citenamefont {Lanzetta}\ \emph {et~al.}(2025)\citenamefont
  {Lanzetta}, \citenamefont {Liu},\ and\ \citenamefont
  {Metlitski}}]{Lanzetta:2025xfw}%
  \BibitemOpen
  \bibfield  {author} {\bibinfo {author} {\bibfnamefont {R.~A.}\ \bibnamefont
  {Lanzetta}}, \bibinfo {author} {\bibfnamefont {S.}~\bibnamefont {Liu}},\ and\
  \bibinfo {author} {\bibfnamefont {M.~A.}\ \bibnamefont {Metlitski}},\
  }\href@noop {} {\bibinfo {title} {{The beginning of the endpoint bootstrap
  for conformal line defects}}} (\bibinfo {year} {2025}),\ \Eprint
  {https://arxiv.org/abs/2508.14964} {arXiv:2508.14964 [cond-mat.str-el]}
  \BibitemShut {NoStop}%
\bibitem [{\citenamefont {Cheng}\ and\ \citenamefont
  {Seiberg}(2026)}]{cheng2026proliferation}%
  \BibitemOpen
  \bibfield  {author} {\bibinfo {author} {\bibfnamefont {M.}~\bibnamefont
  {Cheng}}\ and\ \bibinfo {author} {\bibfnamefont {N.}~\bibnamefont
  {Seiberg}},\ }\bibfield  {title} {\bibinfo {title} {{Proliferation
  transitions from a topological phase in 2+1 dimensions}},\ }\href
  {https://doi.org/10.1103/bfzc-7lm2} {\bibfield  {journal} {\bibinfo
  {journal} {Phys. Rev. B}\ }\textbf {\bibinfo {volume} {114}},\ \bibinfo
  {pages} {055103} (\bibinfo {year} {2026})},\ \Eprint
  {https://arxiv.org/abs/2603.00245} {arXiv:2603.00245 [cond-mat.str-el]}
  \BibitemShut {NoStop}%
\bibitem [{\citenamefont {Barkeshli}\ \emph {et~al.}(2019)\citenamefont
  {Barkeshli}, \citenamefont {Bonderson}, \citenamefont {Cheng},\ and\
  \citenamefont {Wang}}]{Barkeshli:2014cna}%
  \BibitemOpen
  \bibfield  {author} {\bibinfo {author} {\bibfnamefont {M.}~\bibnamefont
  {Barkeshli}}, \bibinfo {author} {\bibfnamefont {P.}~\bibnamefont
  {Bonderson}}, \bibinfo {author} {\bibfnamefont {M.}~\bibnamefont {Cheng}},\
  and\ \bibinfo {author} {\bibfnamefont {Z.}~\bibnamefont {Wang}},\ }\bibfield
  {title} {\bibinfo {title} {Symmetry fractionalization, defects, and gauging
  of topological phases},\ }\href {https://doi.org/10.1103/PhysRevB.100.115147}
  {\bibfield  {journal} {\bibinfo  {journal} {Phys. Rev. B}\ }\textbf {\bibinfo
  {volume} {100}},\ \bibinfo {pages} {115147} (\bibinfo {year} {2019})},\
  \Eprint {https://arxiv.org/abs/1410.4540} {arXiv:1410.4540 [cond-mat.str-el]}
  \BibitemShut {NoStop}%
\bibitem [{\citenamefont {Burnell}(2018)}]{burnell2018anyon}%
  \BibitemOpen
  \bibfield  {author} {\bibinfo {author} {\bibfnamefont {F.~J.}\ \bibnamefont
  {Burnell}},\ }\bibfield  {title} {\bibinfo {title} {Anyon condensation and
  its applications},\ }\href
  {https://doi.org/10.1146/annurev-conmatphys-033117-054154} {\bibfield
  {journal} {\bibinfo  {journal} {Ann. Rev. Condens. Matter Phys.}\ }\textbf
  {\bibinfo {volume} {9}},\ \bibinfo {pages} {307} (\bibinfo {year} {2018})},\
  \Eprint {https://arxiv.org/abs/1706.04940} {arXiv:1706.04940
  [cond-mat.str-el]} \BibitemShut {NoStop}%
\bibitem [{\citenamefont {Drinfeld}\ \emph {et~al.}(2009)\citenamefont
  {Drinfeld}, \citenamefont {Gelaki}, \citenamefont {Nikshych},\ and\
  \citenamefont {Ostrik}}]{2009arXiv0906.0620D}%
  \BibitemOpen
  \bibfield  {author} {\bibinfo {author} {\bibfnamefont {V.}~\bibnamefont
  {Drinfeld}}, \bibinfo {author} {\bibfnamefont {S.}~\bibnamefont {Gelaki}},
  \bibinfo {author} {\bibfnamefont {D.}~\bibnamefont {Nikshych}},\ and\
  \bibinfo {author} {\bibfnamefont {V.}~\bibnamefont {Ostrik}},\ }\href@noop {}
  {\bibinfo {title} {On braided fusion categories {I}}} (\bibinfo {year}
  {2009}),\ \Eprint {https://arxiv.org/abs/0906.0620} {arXiv:0906.0620
  [math.QA]} \BibitemShut {NoStop}%
\bibitem [{\citenamefont {Fidkowski}\ \emph {et~al.}(2013)\citenamefont
  {Fidkowski}, \citenamefont {Chen},\ and\ \citenamefont
  {Vishwanath}}]{PhysRevX.3.041016}%
  \BibitemOpen
  \bibfield  {author} {\bibinfo {author} {\bibfnamefont {L.}~\bibnamefont
  {Fidkowski}}, \bibinfo {author} {\bibfnamefont {X.}~\bibnamefont {Chen}},\
  and\ \bibinfo {author} {\bibfnamefont {A.}~\bibnamefont {Vishwanath}},\
  }\bibfield  {title} {\bibinfo {title} {Non-abelian topological order on the
  surface of a {3D} topological superconductor from an exactly solved model},\
  }\href {https://doi.org/10.1103/PhysRevX.3.041016} {\bibfield  {journal}
  {\bibinfo  {journal} {Phys. Rev. X}\ }\textbf {\bibinfo {volume} {3}},\
  \bibinfo {pages} {041016} (\bibinfo {year} {2013})},\ \Eprint
  {https://arxiv.org/abs/1305.5851} {arXiv:1305.5851 [cond-mat.str-el]}
  \BibitemShut {NoStop}%
\bibitem [{\citenamefont {{Bonati}}\ \emph {et~al.}(2022)\citenamefont
  {{Bonati}}, \citenamefont {{Pelissetto}},\ and\ \citenamefont
  {{Vicari}}}]{BonatiPelissettoVicari}%
  \BibitemOpen
  \bibfield  {author} {\bibinfo {author} {\bibfnamefont {C.}~\bibnamefont
  {{Bonati}}}, \bibinfo {author} {\bibfnamefont {A.}~\bibnamefont
  {{Pelissetto}}},\ and\ \bibinfo {author} {\bibfnamefont {E.}~\bibnamefont
  {{Vicari}}},\ }\bibfield  {title} {\bibinfo {title} {Multicritical point of
  the three-dimensional ${Z}_{2}$ gauge {Higgs} model},\ }\href
  {https://doi.org/10.1103/PhysRevB.105.165138} {\bibfield  {journal} {\bibinfo
   {journal} {\prb}\ }\textbf {\bibinfo {volume} {105}},\ \bibinfo {eid}
  {165138} (\bibinfo {year} {2022})},\ \Eprint
  {https://arxiv.org/abs/2112.01824} {arXiv:2112.01824 [cond-mat.stat-mech]}
  \BibitemShut {NoStop}%
\bibitem [{\citenamefont {Shi}\ and\ \citenamefont
  {Chatterjee}(2025)}]{ShiChatterjee2024}%
  \BibitemOpen
  \bibfield  {author} {\bibinfo {author} {\bibfnamefont {Z.~D.}\ \bibnamefont
  {Shi}}\ and\ \bibinfo {author} {\bibfnamefont {A.}~\bibnamefont
  {Chatterjee}},\ }\bibfield  {title} {\bibinfo {title} {Analytic framework for
  self-dual criticality in $\mathbb{Z}_k$ gauge theory with matter},\ }\href
  {https://doi.org/10.1103/9qrw-p5zn} {\bibfield  {journal} {\bibinfo
  {journal} {Phys. Rev. B}\ }\textbf {\bibinfo {volume} {112}},\ \bibinfo
  {pages} {L081111} (\bibinfo {year} {2025})},\ \Eprint
  {https://arxiv.org/abs/2407.07941} {arXiv:2407.07941 [cond-mat.str-el]}
  \BibitemShut {NoStop}%
\bibitem [{\citenamefont {Serna}\ \emph {et~al.}(2024)\citenamefont {Serna},
  \citenamefont {Somoza},\ and\ \citenamefont {Nahum}}]{Serna:2024fqo}%
  \BibitemOpen
  \bibfield  {author} {\bibinfo {author} {\bibfnamefont {P.}~\bibnamefont
  {Serna}}, \bibinfo {author} {\bibfnamefont {A.~M.}\ \bibnamefont {Somoza}},\
  and\ \bibinfo {author} {\bibfnamefont {A.}~\bibnamefont {Nahum}},\ }\bibfield
   {title} {\bibinfo {title} {{Worldsheet patching, 1-form symmetries, and
  Landau* phase transitions}},\ }\href
  {https://doi.org/10.1103/PhysRevB.110.115102} {\bibfield  {journal} {\bibinfo
   {journal} {Phys. Rev. B}\ }\textbf {\bibinfo {volume} {110}},\ \bibinfo
  {pages} {115102} (\bibinfo {year} {2024})},\ \Eprint
  {https://arxiv.org/abs/2403.04025} {arXiv:2403.04025 [cond-mat.str-el]}
  \BibitemShut {NoStop}%
\bibitem [{\citenamefont {{Dijkgraaf}}\ and\ \citenamefont
  {{Witten}}(1990)}]{DijkgraafWitten}%
  \BibitemOpen
  \bibfield  {author} {\bibinfo {author} {\bibfnamefont {R.}~\bibnamefont
  {{Dijkgraaf}}}\ and\ \bibinfo {author} {\bibfnamefont {E.}~\bibnamefont
  {{Witten}}},\ }\bibfield  {title} {\bibinfo {title} {{Topological gauge
  theories and group cohomology}},\ }\href {https://doi.org/10.1007/BF02096988}
  {\bibfield  {journal} {\bibinfo  {journal} {Commun. Math. Phys.}\ }\textbf
  {\bibinfo {volume} {129}},\ \bibinfo {pages} {393} (\bibinfo {year}
  {1990})}\BibitemShut {NoStop}%
\bibitem [{\citenamefont {Dumitrescu}\ \emph {et~al.}(2024)\citenamefont
  {Dumitrescu}, \citenamefont {Niro},\ and\ \citenamefont
  {Thorngren}}]{Dumitrescu:2024jko}%
  \BibitemOpen
  \bibfield  {author} {\bibinfo {author} {\bibfnamefont {T.~T.}\ \bibnamefont
  {Dumitrescu}}, \bibinfo {author} {\bibfnamefont {P.}~\bibnamefont {Niro}},\
  and\ \bibinfo {author} {\bibfnamefont {R.}~\bibnamefont {Thorngren}},\
  }\href@noop {} {\bibinfo {title} {{Symmetry Breaking from Monopole
  Condensation in QED$_3$}}} (\bibinfo {year} {2024}),\ \Eprint
  {https://arxiv.org/abs/2410.05366} {arXiv:2410.05366 [hep-th]} \BibitemShut
  {NoStop}%
\bibitem [{\citenamefont {{Seiberg}}\ and\ \citenamefont
  {{Witten}}(2016)}]{SeibergWitten2016}%
  \BibitemOpen
  \bibfield  {author} {\bibinfo {author} {\bibfnamefont {N.}~\bibnamefont
  {{Seiberg}}}\ and\ \bibinfo {author} {\bibfnamefont {E.}~\bibnamefont
  {{Witten}}},\ }\bibfield  {title} {\bibinfo {title} {{Gapped boundary phases
  of topological insulators via weak coupling}},\ }\href
  {https://doi.org/10.1093/ptep/ptw083} {\bibfield  {journal} {\bibinfo
  {journal} {Prog. Theor. Exp. Phys.}\ }\textbf {\bibinfo {volume} {2016}},\
  \bibinfo {eid} {12C101} (\bibinfo {year} {2016})},\ \Eprint
  {https://arxiv.org/abs/1602.04251} {arXiv:1602.04251 [cond-mat.str-el]}
  \BibitemShut {NoStop}%
\bibitem [{\citenamefont {{Seiberg}}\ \emph {et~al.}(2016)\citenamefont
  {{Seiberg}}, \citenamefont {{Senthil}}, \citenamefont {{Wang}},\ and\
  \citenamefont {{Witten}}}]{DualityWeb}%
  \BibitemOpen
  \bibfield  {author} {\bibinfo {author} {\bibfnamefont {N.}~\bibnamefont
  {{Seiberg}}}, \bibinfo {author} {\bibfnamefont {T.}~\bibnamefont
  {{Senthil}}}, \bibinfo {author} {\bibfnamefont {C.}~\bibnamefont {{Wang}}},\
  and\ \bibinfo {author} {\bibfnamefont {E.}~\bibnamefont {{Witten}}},\
  }\bibfield  {title} {\bibinfo {title} {{A duality web in 2 + 1 dimensions and
  condensed matter physics}},\ }\href
  {https://doi.org/10.1016/j.aop.2016.08.007} {\bibfield  {journal} {\bibinfo
  {journal} {Ann. Phys.}\ }\textbf {\bibinfo {volume} {374}},\ \bibinfo {pages}
  {395} (\bibinfo {year} {2016})},\ \Eprint {https://arxiv.org/abs/1606.01989}
  {arXiv:1606.01989 [hep-th]} \BibitemShut {NoStop}%
\bibitem [{\citenamefont {Delmastro}\ and\ \citenamefont
  {Gomis}()}]{GomisQuantumSymmetry}%
  \BibitemOpen
  \bibfield  {author} {\bibinfo {author} {\bibfnamefont {D.}~\bibnamefont
  {Delmastro}}\ and\ \bibinfo {author} {\bibfnamefont {J.}~\bibnamefont
  {Gomis}},\ }\bibfield  {title} {\bibinfo {title} {Symmetries of {Abelian}
  {Chern-Simons} theories and arithmetic},\ }\href
  {https://doi.org/10.1007/JHEP03(2021)006} {\bibfield  {journal} {\bibinfo
  {journal} {JHEP}\ }\textbf {\bibinfo {volume} {03}}\bibfield  {number}
  {\bibinfo  {number} { (2021)},\ \bibinfo {pages} {006}},\ }\Eprint
  {https://arxiv.org/abs/1904.12884} {arXiv:1904.12884 [hep-th]} \BibitemShut
  {NoStop}%
\bibitem [{\citenamefont {{Levin}}\ and\ \citenamefont
  {{Stern}}(2012)}]{LevinStern2012}%
  \BibitemOpen
  \bibfield  {author} {\bibinfo {author} {\bibfnamefont {M.}~\bibnamefont
  {{Levin}}}\ and\ \bibinfo {author} {\bibfnamefont {A.}~\bibnamefont
  {{Stern}}},\ }\bibfield  {title} {\bibinfo {title} {{Classification and
  analysis of two-dimensional Abelian fractional topological insulators}},\
  }\href {https://doi.org/10.1103/PhysRevB.86.115131} {\bibfield  {journal}
  {\bibinfo  {journal} {\prb}\ }\textbf {\bibinfo {volume} {86}},\ \bibinfo
  {eid} {115131} (\bibinfo {year} {2012})},\ \Eprint
  {https://arxiv.org/abs/1205.1244} {arXiv:1205.1244 [cond-mat.str-el]}
  \BibitemShut {NoStop}%
\end{thebibliography}%

\section*{End Matter}

\section{More on anyon condensation and anyon permuting symmetry}

We now review some known facts about anyon condensation in a physically intuitive language. A more comprehensive treatment can be found in Ref.~\cite{bais2009condensate, Barkeshli:2014cna,burnell2018anyon}. We start from a topological order $\mathcal{C}$ -- cases of our interest include the double Ising $\{1,\sigma,\psi\}\times\{1,\bar{\sigma},\bar{\psi}\}$ and $SU(2)_k\times SU(2)_{-k}$. We consider situations where there is an abelian bosonic anyon $J$ that is order-$2$, meaning $J\times J= 1$. For the double Ising $J=\psi\bar{\psi}$, and for the $SU(2)_k\times SU(2)_{-k}$ theory $J$ is the $(k/2,k/2)$ anyon.

We now condense the abelian boson $J$ to obtain a new topological order $\mathcal{C}'$, since in cases we focus here, $1+J$ forms a condensable algebra. Now $J$ is created by a local operator, call it $X_J$. Since $J$ is order-$2$, we can choose the local operator so that $X_J^2=1$. The remaining topological order $\mathcal{C}'$ can have a $\Z_2$ global zero-form symmetry $g$, and the particle creation operator $X_J$ will be charged under $g$: $g^{\dagger}X_Jg=-X_J$. In addition, the global $\mathbb{Z}_2$ symmetry acts non-trivially as an exchange symmetry on the anyons in $\mathcal{C}'$, which we will elaborate below. If we now gauge this global $\Z_2$, we go back to the original topological order $\mathcal{C}$, so condensing $J$ and gauging $g$ should be viewed as inverse operations of each other \cite{2009arXiv0906.0620D,bais2009condensate,Barkeshli:2014cna}.

Now following the standard rule of anyon condensation, an anyon $a$ in $\mathcal{C}$ will be confined (deleted from the excitation spectrum) if $\theta_{a,J}^{a'}=\frac{\theta_{a'}}{\theta_a \theta_J}\neq 1$ for any fusion outcome $a'$ in $a\times J$, i.e. $a$ has nontrivial mutual braiding statistical angle with $J$ \cite{bais2009condensate,burnell2018anyon}. And all anyons in $\mathcal{C}$ that have trivial mutual statistics with $J$ remain as topological excitations in $\mathcal{C}'$. Furthermore, in $\mathcal{C}'$, we should also identify $a$ with $a\times J$, since the two only differ by a local operator. 

There is a special situation if there is an anyon $X$ in $\mathcal{C}$, called the ``fixed point anyon,'' that (1) has $\theta_{XJ}=1$ and (2) is invariant under attaching $J$: $X\times J=X$. For the double Ising topological order with $J=\psi\bar{\psi}$, the fixed point anyon is $X=\sigma\bar{\sigma}$. For $SU(2)_k\times SU(2)_{-k}$ with $J=(k/2,k/2)$, the fixed point anyon is $X=(k/4,k/4)$ if $k$ is even, and for odd $k$ there is no fixed point anyon. For the fixed point anyon $X$, attaching $J$ can only rearrange the internal degrees of freedom of the non-abelian anyon $X$. This remains true in $\mathcal{C}'$, where $X$ survives as a nontrivial anyon. In $\mathcal{C}'$, $J$ is created by a local $\Z_2$ operator $X_J$, so we can label the internal degree of freedom on $X$ through $\pm1$ eigenvalues of $X_J$. We can write 
\begin{equation}
    X\rightarrow X_+\oplus X_-,\hspace{5pt} X_{\pm}=\frac{1\pm X_J}{2}X,
\end{equation}
where $X_{\pm}$ is the projection of $X$ to the $X_J=\pm1$ sector. 

Two important observations follow:
\begin{enumerate}
    \item Since all the remaining anyons in $\mathcal{C}'$ are mutually local with $J$, there is no anyon braiding process that mixes $X_+$ and $X_-$ -- they are genuinely different anyons now (instead of being part of the same non-Abelian anyon).
    \item Under the global $\Z_2$ symmetry $g$, $X_+\leftrightarrow X_-$. This also means that both $X_{\pm}$ must be nontrivial, with identical quantum dimension and topological spin.
\end{enumerate}

\textbf{\boldmath The $k=2$ case.} We can illustrate the above points quite simply using the double Ising example with $J=\psi\bar{\psi}$ and $X=\sigma\bar{\sigma}$. Denote the two Majorana zero modes associated with $\sigma$ and $\bar{\sigma}$ as $\gamma$ and $\bar{\gamma}$. The $\sigma\bar{\sigma}$ anyon has quantum dimension $\sqrt{2}\times \sqrt{2}=2$, and these two states are distinguished by the parity of $i\gamma\bar{\gamma}$, which is now a local operator -- exactly the $X_J$ in previous discussion. For each eigenvalue $i\gamma\bar{\gamma}=\pm1$, we have a unique anyons state -- these are nothing but the abelian anyons $e$ and $m$. Since $X_J$ is charged under the global $\Z_2$, the global $\Z_2$ symmetry flips the eigenvalue of $X_J$, hence exchanges $e$ and $m$. 

\textbf{\boldmath The $k=6$ case.} Here, we condense $J=(3,3)$ in $SU(2)_6\times SU(2)_{-6}$, the resulting topological order, denoted as $SO(4)_{6,-6}$, is a modular extension of $SO(3)_6\times SO(3)_{-6}$, where $SO(3)_6$ is a non-modular tensor category, since it contains a transparent fermion.\cite{PhysRevX.3.041016} We list the $14$ anyons of $SO(4)_{6,-6}$, and their counterparts before the condensation in Table \ref{tab:k=6}. In particular, the anyon $X=\left(\frac{3}{2},\frac{3}{2}\right)$ splits into two anyons $\rho_e$ and $\rho_m$ in $SO(4)_{6,-6}$. The two are self-bosons, and satisfy the following fusion rules. 
\begin{align}
\rho_e\times \rho_e&=\rho_m\times \rho_m=\mathbf{1}+(1,\tilde{s})+(\tilde{s},1)+(s,s),\nonumber\\
\rho_e \times \rho_m &=f+ (1,s)+(s,1)+ (s,\tilde{s}),\\
\rho_e \times f &= \rho_m,~~\rho_m\times f = \rho_e.\nonumber
\end{align}

The two anyons $\rho_e$ and $\rho_m$ have non-trivial mutual braiding statistics, with $\theta_{\rho_e,\rho_m}^{\rho_{e}\times \rho_m}\in \{-1,i,-i\}$. Under the global $\mathbb{Z}_2$ symmetry, $\rho_e$ and $\rho_m$ are exchanged. 

\vspace*{0pt}
\begin{table}[H]
\centering
\renewcommand{\arraystretch}{1.2}
\begin{tabular}{ |c | c | c | c |}
\hline
$(J_L,J_R)\in \mathcal{C}$ & $h\!\!\mod 1$ & q.d. & anyons in $\mathcal{C'}$ \\
\hline
$(0,0)\sim (3,3)$ & $0$ & 1 & $(1,1)\sim (f,f)$ \\
\hline
$(1,0)\sim (2,3)$ & $\frac{1}{4}$ & $1+\sqrt{2}$ & $(s,1)\sim (\tilde s,f)$ \\
\hline
$(2,0)\sim (1,3)$ & $\frac{3}{4}$ & $1+\sqrt{2}$ & $(\tilde s,1)\sim (s,f)$ \\
\hline
$(3,0)\sim (0,3)$ & $\frac{1}{2}$ & 1 & $(f,1)\sim (1,f)$ \\
\hline
$(0,1)\sim (3,2)$ & $-\frac{1}{4}$ & $1+\sqrt{2}$ & $(1,s)\sim (f,\tilde s)$ \\
\hline
$(1,1)\sim (2,2)$ & $0$ & $3+2\sqrt{2}$ & $(s,s)\sim (\tilde s,\tilde s)$ \\
\hline
$(2,1)\sim (1,2)$ & $\frac{1}{2}$ & $3+2\sqrt{2}$ & $(\tilde s,s)\sim (s,\tilde s)$ \\
\hline
$(3,1)\sim (0,2)$ & $-\frac{3}{4}$ & $1+\sqrt{2}$ & $(f,s)\sim (1,\tilde s)$ \\
\hline
$\left(\frac{1}{2},\frac{1}{2}\right)\sim \left(\frac{5}{2},\frac{5}{2}\right)$ & $0$ & $2+\sqrt{2}$ & $\mu$ \\
\hline
$\left(\frac{3}{2},\frac{1}{2}\right)\sim \left(\frac{3}{2},\frac{5}{2}\right)$ & $\frac{3}{8}$ & $2+2\sqrt{2}$ & $\nu$ \\
\hline
$\left(\frac{5}{2},\frac{1}{2}\right)\sim \left(\frac{1}{2},\frac{5}{2}\right)$ & $0$ & $2+\sqrt{2}$ & $\mu'$ \\
\hline
$\left(\frac{1}{2},\frac{3}{2}\right)\sim \left(\frac{5}{2},\frac{3}{2}\right)$ & $-\frac{3}{8}$ & $2+2\sqrt{2}$ & $\nu'$ \\
\hline
$\left(\frac{3}{2},\frac{3}{2}\right)_+$ & $0$ & $2+\sqrt{2}$ & $\rho_e$ \\
\hline
$\left(\frac{3}{2},\frac{3}{2}\right)_-$ & $0$ & $2+\sqrt{2}$ & $\rho_m$ \\
\hline
\end{tabular}
\caption{Condensation of $J=(3,3)$ in topological order $\mathcal{C}=SU(2)_6 \times SU(2)_{-6}$ results in topological order $\mathcal{C'}=SO(4)_{6,-6}$. The topological spin of anyon $a$ is given by $\theta_a = e^{i 2\pi h_a}$. And q.d. stands for quantum dimension. Any anyon with half-integral $J_L$ and $J_R$ (i.e., the last $6$ anyons in the table) has semionic mutual statistics $\theta_{a,f}=-1$ with $f\equiv (f,1)\sim (1,f)$.}
\label{tab:k=6}
\end{table}

\textbf{Higher-$k$ cases.} Starting from k=6, the topological phase is no longer given by a finite-group gauge theory. The symmetry permutes two self-bosonic anyons. Except for special values of $k$ (e.g., $k =10, 26, 28$ for $k<30$), the permuted anyons always have non-trivial mutual braiding. In this regard, the symmetry might be considered as EM exchange symmetry in a loose sense. Nevertheless, when k = 10, 26 or 28, there are fusion channels of the two self-bosonic anyons that braid trivially. In such cases, we can only say that the symmetry always permutes two self-bosonic anyons that \textit{can} braid non-trivially, i.e., there always exist fusion channels that the two braid non-trivially.


\clearpage
\onecolumngrid

\begin{center}
\textbf{\large Supplementary Material for\\
``Self-dual Higgs transitions: Toric code and beyond''}
\end{center}

\section{Previous attempts on field-theoretic understanding}

We now review two previous attempts to understand the self-dual transition of the toric code: the gauged $XY$ universality proposal \cite{BonatiPelissettoVicari} and the abelian double Chern-Simons proposal \cite{ShiChatterjee2024,Serna:2024fqo}. We will explain the difficulties encountered by each proposal.

\subsection{The gauged $XY$ proposal}

In the gauged $XY$ proposal, the $e$ and $m$ particles are first treated as decoupled real scalars $\phi_{e}$ and $\phi_m$:
\begin{equation}
    \mathcal{L}[\phi_{e},\phi_m]=\mathcal{L}_{\rm Ising}[\phi_e]+\mathcal{L}_{\rm Ising}[\phi_m],
\end{equation}
where $\mathcal{L}_{\rm Ising}$ is the standard Ising-$\phi^4$ field theory Lagrangian. Then we turn on a symmetry-allowed perturbation $\phi_e^2\phi_m^2$, which is known to drive the doubled Ising theory to the $XY$ universality class, with $\phi_e$ and $\phi_m$ being the real and imaginary parts of a complex $XY$ scalar. 

The key difficulty with this proposal is reproducing the correct nonlocal properties of the $e$ and $m$ particles, namely their mutual $\pi$-statistics in the toric code phase when $\phi_{e,m}$ are gapped and uncondensed. A natural attempt is to couple $\phi_e$ and $\phi_m$ to dynamical $\mathbb{Z}_2$ gauge fields. However, this approach faces a fundamental obstruction: coupling both fields to the same $\mathbb{Z}_2$ gauge field (also known as $XY^*$) fails to reproduce the correct mutual statistics, while coupling them to two distinct $\mathbb{Z}_2$ gauge fields (also known as $XY^{**}$), possibly with additional topological terms~\cite{DijkgraafWitten}, leads instead to a $\mathbb{Z}_2 \times \mathbb{Z}_2$ topological order rather than the $\mathbb{Z}_2$ topological order as in the toric code.

\subsection{The abelian double Chern-Simons proposal}

In the double Chern-Simons proposal, the toric code topological order is represented using a $U(1)\times U(1)$ mutual Chern-Simons (also known as BF) theory. More explicitly, we have two $U(1)$ gauge fields $a_1$ and $a_2$, with the Lagrangian given by the BF term:
\begin{equation}
\label{eq:BF}
    \mathcal{L}_{BF}=i\frac{k}{2\pi} a_1\wedge da_2,
\end{equation}
where $k=2$ represents the standard toric code, and a general integer $k$ represents a $\Z_k$ gauge theory. We then introduce two complex scalars $\phi_{1,2}$ that couple to $a_{1,2}$ gauge fields, respectively:
\begin{eqnarray}  \mathcal{L}&=&\sum_{i=1,2}\left(|D_{a_i}\phi_i|^2+r|\phi_i|^2+\lambda|\phi_i|^4\right)+\lambda'|\phi_1|^2|\phi_2|^2 \nonumber \\
& &+\sum_{i=1,2}\frac{1}{4e^2}(\partial\times a_i)^2+i\frac{k}{2\pi} a_1\wedge da_2.
\label{eq:BFHiggs}
\end{eqnarray}

The topologically ordered phase corresponds to $r>0$, where the fields $\phi_{1,2}$ are gapped and, through flux attachment from the Chern-Simons term, create the gapped $e$ and $m$ excitations. The self-duality symmetry $\mathbb{Z}_2^{SD}$ simply exchanges $\phi_1 \leftrightarrow \phi_2$ and $a_1 \leftrightarrow a_2$. For $r<0$, the scalar fields condense and the topological order is destroyed. If $\lambda'$ is large and positive, the Higgs phase may spontaneously break $\mathbb{Z}_2^{SD}$ by condensing only $\phi_1$ or only $\phi_2$.

The abelian BF theory proposal Eq.~\eqref{eq:BFHiggs} can be a perfectly consistent theory describing a self-dual transition of $\Z_k$ toric code for large $k$ (for example $k\geq 4$) \cite{ShiChatterjee2024}. However, for $k=2$ (the toric code) there are several notable difficulties: 
\begin{enumerate}
    \item The theory for a direct continuous transition is analytically controlled, via large-$N$ expansion, only in the small $\lambda'$ regime, where the Higgs phase does not spontaneously break the $\Z_2^{SD}$ symmetry.
    \item The monopole operators of the gauge fields are allowed as local perturbations, and for $k=2$ these monopoles are most likely relevant \cite{ShiChatterjee2024}. These monopole perturbations drive the theory into a strong-coupling regime, making it difficult to draw any further conclusions. For a recent study addressing related issues from a field-theoretic point of view, see \cite{Dumitrescu:2024jko}.
    \item There is an additional symmetry of the microscopic theory, namely the combination of time-reversal and $\Z_2^{SD}$, that is not correctly represented in the BF theory. Since this point has not been discussed in previous literature, we explain it in more detail in the following. 
\end{enumerate}

\subsubsection{Symmetry of toric code and field theory}

The toric code, as an untwisted $\Z_2$ topological order, admits global $\Z_2\times\Z_2^T$ symmetry, where the unitary $\Z_2$ is the ``self-duality'' exchanging $e\leftrightarrow m$, and time-reversal symmetry $\mathcal{T}$ acts trivially on the anyons. Combining the two symmetry operations gives another time-reversal $\tilde{\mathcal{T}}: e\leftrightarrow m, i\to-i$. This $\Z_2^{\tilde{T}}$ symmetry turns out to be quite subtle in the standard field theory representation.

To implement a time-reversal operation that exchanges the $e$ ($a_1$-charge) and $m$ ($a_2$-charge) excitations , while keeping the mutual CS action Eq.~\eqref{eq:BF} invariant, the only way is to have
\begin{equation}
    \tilde{\mathcal{T}}: a_1\to \mp a_2, a_2\to \pm a_1,
    \label{eq:TKmatrix}
\end{equation}
where ``$a_{1,2}$'' above represent the time-components. The space components undergo similar transforms but with the opposite sign. The problem is that $\tilde{\mathcal{T}}$ above does not generate a $\Z_2$ operation: $\tilde{\mathcal{T}}^2=\mathcal{C}$ where $\mathcal{C}$ is the charge conjugation $a_{1,2}\to -a_{1,2}$. In other words, in the abelian CS formulation of the toric code phase, the combination of the self-duality and time-reversal symmetry can only be $\Z_4$, not $\Z_2$. {More generally, the $K$-matrix formalism for abelian topological orders is inadequate to describe time-reversal symmetry that exchanges anyons with non-trivial mutural statistics \cite{PhysRevX.3.041016}}.

Deep in the topologically ordered phase, the above issue is not very serious \cite{SeibergWitten2016,DualityWeb,GomisQuantumSymmetry}: when the theory is put on-shell, the Wilson lines $\int a_{1,2}$ are either $0$ or $\pi$, so we can identify $a_{1,2}\sim -a_{1,2}$, and the $e$-$m$ exchanging time-reversal symmetry becomes effectively $\Z_2$. The need to be on-shell for the symmetry to become manifest does make certain aspects difficult. For example, it is not clear how to put the mutual CS theory on a non-orientable manifold twisted by $\tilde{\mathcal{T}}$. It is also difficult to study symmetric edge theories \cite{LevinStern2012} in this formulation.

The issue becomes serious if we consider phase transitions, for example by introducing critical scalar fields coupled to $a_{1,2}$ as in Eq.~\eqref{eq:BFHiggs}. Now the charge conjugation symmetry acts nontrivially in the IR, so we should treat the symmetry honestly as $\Z_4$, not $\Z_2$. This makes the abelian CS formulation unsuitable for describing the transition of the toric code enriched by $\mathbb{Z}_2^{\tilde{\mathcal{T}}}$ symmetry that is realized in the microscopic theory. Within the BF formulation, the only possible way to resolve this issue is to assume that the monopole perturbation drives the theory to a new fixed point, at which the charge conjugation symmetry acts trivially on the IR theory. While this scenario is possible, it appears difficult to either justify or falsify.

The $SO(4)_{2,-2}$ formulation of the toric code does not have the above issue. Time-reversal symmetry acts as the improper $\Z_2\in O(4)$ transform on the $4$-component scalar $\Phi$, and the $e\leftrightarrow m$ self-duality symmetry acts as the $\Z_2$ flux conservation. The combination of the two still squares to identity -- perhaps the easiest way to see this is to notice that there is no other global symmetry manifest in the field theory Eq.~\eqref{eq:2,-2Higgs}. The symmetry is manifestly correct at the Lagrangian level, which is another reason to favor the $SO(4)_{2,-2}$ CSH theory when describing the self-dual transition of the toric code. 


\end{document}